\documentclass{article}
\usepackage {appendix}
\usepackage{algorithm}
\usepackage{algorithmic}
\usepackage{amsthm}
\usepackage{PRIMEarxiv}
\usepackage{amsmath}
\usepackage[utf8]{inputenc} % allow utf- input
\usepackage[T1]{fontenc}    % use 8-bit T1 fonts
\usepackage{hyperref}       % hyperlinks
\usepackage{url}            % simple URL typesetting
\usepackage{booktabs}       % professional-quality tables
\usepackage{amsfonts}       % blackboard math symbols
\usepackage{nicefrac}       % compact symbols for 1/2, etc.
\usepackage{microtype}      % microtypography
\usepackage{lipsum}
\usepackage{fancyhdr}       % header
\usepackage{graphicx}       % graphics
\usepackage{textcomp}
\usepackage{subfig}

\graphicspath{{media/}}     % organize your images and other figures under media/ folder

\title{Quasi-binary encoding based quantum alternating operator ansatz
}

\author{
  Bingren Chen, Hanqing Wu, Haomu Yuan, Lei Wu, Xin Li \\
  CCB fintech \\
  \texttt{bingren@mail.ustc.edu.cn} \\
}

\begin{document}
\maketitle

\begin{abstract}
  This paper proposes a quasi-binary encoding based algorithm for solving a specific quadratic optimization models with discrete variables, in the quantum approximate optimization algorithm (QAOA) framework. The quadratic optimization model has three constraints: 1. Discrete constraint, the variables are required to be integers. 2. Bound constraint, each variable is required to be greater than or equal to an integer and less than or equal to another integer. 3. Sum constraint, the sum of all variables should be a given integer.

  To solve this optimization model, we use quasi-binary encoding to encode the variables. For an integer variable with upper bound $U_i$ and lower bound $L_i$, this encoding method can use at most $2\log_2 (U_i-L_i+1)$ qubits to encode the variable. Moreover, we design a mixing operator specifically for this encoding to satisfy the hard constraint model. In the hard constraint model, the quantum state always satisfies the constraints during the evolution, and no penalty term is needed in the objective function. In other parts of the QAOA framework, we also incorporate ideas such as CVaR-QAOA and parameter scheduling methods into our QAOA algorithm.
  
  In the financial field, by introducing precision, portfolio optimization problems can be reduced to the above model. We will use portfolio optimization cases for numerical simulation. We design an iterative method to solve the problem of coarse precision caused by insufficient qubits of the simulators or quantum computers. This iterative method can refine the precision by multiple few-qubit experiments.
  \end{abstract}

\keywords{Quantum approximate optimization algorithm \and Quantum alternative operator Ansatz \and Portfolio optimization}
\section{Introduction}

Quantum computing is an emerging field that aims to harness the power of quantum phenomena to perform computations beyond the capabilities of classical computers. The field traces its origins to the 1980s, when Richard Feynman suggested the possibility of using quantum systems to simulate complex physical processes \cite{feynman_quantum_1985}. Since then, quantum computing has become an interdisciplinary field that involves physics, computer science, chemistry, biology, etc \cite{schuld_introduction_2015, dusek_quantum_2006, cao_quantum_2019, levine_quantum_2009, lambert_quantum_2013}.

A near-term major challenge in quantum computing is to develop algorithms that can run on noisy intermediate-scale quantum (NISQ) devices. Quantum heuristics algorithms, including quantum approximate optimization algorithm (QAOA) \cite{farhi_quantum_2014}, variational quantum eigensolver (VQE) \cite{tilly_variational_2022} and  quantum annealing (QA) \cite{finnila_quantum_1994}, are a type of algorithms that are tailored for NISQ devices. These algorithms are designed to be resilient to noise and other sources of error in the quantum hardware, making them suitable for NISQ devices.

As one of the most prominent quantum heuristics algorithms, QAOA is designed to solve combinatorial optimization problems. The algorithm has demonstrated its effectiveness on various problems, such as MaxCut \cite{farhi_quantum_2014}, MaxSat \cite{hadfield_quantum_2019}, and graph partitioning \cite{li_large-scale_2022}. QAOA is expected to play a key role in the development of real-world quantum applications.

QAOA for solving constrained combinatorial optimization can be divided into two models: the soft constraint model and the hard constraint model \cite{hadfield_quantum_2019, hodson_portfolio_2019}. The soft constraint model incorporates the constraints into the objective function as penalty terms, which can be adjusted by a penalty factor. However, choosing an appropriate penalty factor is challenging, as it affects the balance between feasibility and optimality of the solutions \cite{wang_xy_2020, niroula_constrained_2022}. When the penalty factor is large, the model tends to output solutions that satisfy the constraint conditions, while ignoring the differences in the value of the original objective function. On the contrary, when the penalty factor is small, the probability of infeasible solutions output by the model will increase. The hard constraint model, also known as quantum alternating operator ansatz \cite{hadfield_quantum_2019} or QAOAz \cite{slate_quantum_2021} or C-QAOA \cite{ye_towards_2023}, ensures that the quantum state represents a feasible solution by designing operators that control the qubit evolution. This model has a higher chance of finding the optimal solution, but it is more complex to implement.

QAOAz algorithms are widely used to solve combinatorial optimization problems with binary variables, such as Max Cut and Traveling Salesman Problem \cite{hadfield_quantum_2019}. In these problems, variables are only represented by 0 and 1, indicating one of two options. Therefore, QAOAz algorithms do not need a specific encoding method for these problems. However, some more practical optimization models, such as portfolio optimization problems in the financial field \cite{hodson_portfolio_2019, palmer_quantum_2021}, will involve integer encoding to represent the number of shares . One of the main challenges for applying QAOAz to this kind of problem is that the existing encoding methods are inefficient, as the qubits increase linearly to represent more choices, which results in limited scale and precision of the problem solving. For instance, the current state-of-the-art such as Hodson's enoding \cite{hodson_portfolio_2019} or quantum-walk QAOA (QW-QAOA) \cite{slate_quantum_2021} only consider three discrete positions for portfolio optimization: long, short, and no position, rather than the percentage of investment or integer shares. Another challenge is that other efficient encoding methods (such as binary encoding or Palmer encoding \cite{palmer_quantum_2021}) do not accommodate hard constraints, because it is difficult to construct a QAOAz mixer that satisfies the encoding rule. Although some general QAOAz mixer preparation methods \cite{bartschi_grover_2020, fuchs_constraint_2022} have emerged, or there are some improvement schemes for the solution quality of soft constraint methods \cite{hao_exploiting_2022, brandhofer_benchmarking_2022, herman_constrained_2023}, these schemes are difficult to implement effectively on the business side due to efficiency issues or their own assumptions.

This paper will propose a quasi-binary encoding based QAOA (QB-QAOA) that aims to solve a quadratic optimization problem with integer variables in hard constraint way. The constraints of the model involve the sum constraint and bound constraint of integer variables. The optimization model can be written as:

\begin{equation}
  \label{e1}
  \begin{aligned}
  &\min_{\mathbf{x}}\  \sum_{i=1}^n \sum_{j=1}^n \mathbf{\sigma}_{i,j} \mathbf{x}_i \mathbf{x}_j + \sum_{i=1}^n \mathbf{\mu}_i \mathbf{x}_i\\
  &\textbf{s.t.} \sum_{i=1}^n \mathbf{x}_{i} =D, \ L_i \leq \mathbf{x}_i \leq U_i, \   \ \mathbf{x}_i \in \mathbb{Z} \ ,\forall i \in \{1, 2, \cdots, n\}
  \end{aligned}
\end{equation}

where $n$ is the number of the integer variables in the model,  $\mathbf{x}_i$ denotes the $i$-th variable, $L_i$ and $U_i$ denote the lower bound and upper bound of $\mathbf{x}_i$, and $D$ is the sum of all elements in ${\mathbf{x}}$.

To convert the integer variables to decimal variables, we define a new parameter $\alpha$ whose value is equal to $\frac{1}{D}$, which represents the precision coefficient of the variables. By multiplying both sides of the constraints by $\alpha$, we can rewrite some of the parameters and variables as follows:

\begin{equation}
\label{e2}
\begin{aligned}
&\mathbf{w}_i = \mathbf{x}_i \alpha\\
&\mathbf{s}_{i,j} = \frac{\mathbf{\sigma}_{i,j}}{\alpha^2}\\
&\mathbf{e}_i = -\frac{\mathbf{\mu}_i}{\alpha}\\
&l_i = L_i \alpha\\
&u_i = U_i \alpha\\
& 1 = D \alpha
\end{aligned}
\end{equation}

Using these new expressions, we can reformulate Equation \ref{e1} as Equation \ref{e3}:

\begin{equation}
\label{e3}
\begin{aligned}
&\min_{\mathbf{w}}\ \sum_{i=1}^n \sum_{j=1}^n \mathbf{s}_{i,j} \mathbf{w}_i \mathbf{w}_j - \sum_{i=1}^n \mathbf{e}_i \mathbf{w}_i\\
&\textbf{s.t.} \sum_{i=1}^n \mathbf{w}_{i} =1, \ l_i \leq \mathbf{w}_i \leq u_i,\  \ \frac{\mathbf{w}_i}{\alpha} \in \mathbb{Z}, \forall i \in \{1, 2, \cdots, n\}
\end{aligned}
\end{equation}

Equation \ref{e3} is analogous to the Markowitz model \cite{markowitz_foundations_1991} for portfolio optimization, which can assist investors in optimizing their portfolio by trading off risk and return. Here, $\mathbf{s}_{i,j}$ represents the covariance of historical interest rates between asset $i$ and asset $j$, and $\mathbf{e}_{i}$ denotes the expectation of historical interest rate for asset $i$. When we expect to use QAOA to solve the portfolio optimization problem with percentage weights, we pre-determine a parameter coefficient $\alpha$, whose reciprocal is an integer, and then transform Equation \ref{e3} into Equation \ref{e1} to solve.

QB-QAOA requires only $2\log_2 (U_i-L_i+1)$ qubits to represent each variable $\mathbf{x}_i$. Moreover, we propose a novel mixing operator for QB-QAOA, which has a circuit depth ranging from constant to linear depending on the problem structure. In the numerical simulation, we apply QB-QAOA to the portfolio optimization problem, and also introduce CVaR-QAOA \cite{barkoutsos_improving_2020} and parameter scheduling techniques \cite{brandhofer_benchmarking_2022} to enhance the solution quality. We also propose a precision increasing iterative method, which is motivated by the fact that the current real quantum computers or simulators cannot provide enough qubits to perform precise calculations. During the iterative process, our iterative method initially uses coarse precision (large $\alpha$ in Equation \ref{e2}) for operations, before subsequently increasing the precision ($\alpha$ gets smaller) while maintaining the same qubits count.

The paper is organized as follows: In Section \ref{sec3}, we introduce the fundamental concepts the QAOA framework. In Section \ref{sec4}, we present the Quasi-Binary based QAOA algorithm including the design of quasi-binary encoding and mixing operator. In Section \ref{sec5}, we provide a series of numerical simulations to evaluate the performance of QB-QAOA. In Section \ref{sec6}, we present the precision increasing iterative method and corresponding simulation results. In Section \ref{sec7}, a conclusion is presented.

\section{Quantum Approximate Optimization Algorithm Framework}
\label{sec3}

There are four main components in
QAOA framework: encoding, initial states, phase-separation operator, and mixing operator.

\subsection{Encoding}

The \textbf{encoding} is the process of mapping the solutions of a problem to quantum states and reading the outcomes of qubit measurements as solutions. The purpose of encoding is to represent decimal or integer variables in an optimization model using binary variables. The encoding also determines how many qubits are needed for each asset. In a hard constraint model, the encoding also influences the difficulty to design the mixing operator.

The encoding of QAOA varies depending on the application of interest. For example, for the maxcut problem, the encoding is straightforward, with 0 and 1 representing different sets of vertices. For portfolio optimization, the encoding uses the qubits to express integers. Binary and one-hot encoding are often used to present one integer variable. The number of qubits required for the former scales logarithmically with the range of integers represented, while the latter scales linearly.

Besides binary and one-hot encoding, QAOA encoding also includes domain wall encoding \cite{chancellor_domain_2019}, Hodson encoding \cite{hodson_portfolio_2019}, Quantum Walk encoding  \cite{slate_quantum_2021}, Palmer encoding \cite{palmer_quantum_2021}, etc. Domain wall encoding can save one qubit for each discrete variable and only requires linear connections instead of full connections between qubits compared with one-hot encoding \cite{chancellor_domain_2019}. Hodson encoding, Palmer encoding, and Quantum Walk encoding are only suitable for portfolio optimization problems. Hodson encoding uses two qubits to represent the encoding state of each position, where 01 means long position (1), 10 means short position (-1), 00 and 11 mean no position (0) \cite{hodson_portfolio_2019}. When the choices of the variable increase, for example, if $x_i \in \{-2,-1,0,1,2\}$, 4 qubits should be used, such as 0101 represents 2 or 1010 represents -2. Therefore, the number of qubits for each variable will increase linearly as the choice of that variable increases. Quantum Walk encoding can encode across assets and assign a number to each valid portfolio starting from 0, making it the most efficient \cite{hodson_portfolio_2019}. However, since Quantum Walk encoding cannot independently compute the phase separation operator, it still relies on the Hodson encoding circuit. Palmer encoding is specially designed for bound constraints, and its number of qubits is at most one more than binary encoding \cite{palmer_quantum_2021}, but it does not support hard constraint models.

To our knowledge, our quasi-binary encoding is the only encoding in QAOA that satisfies the following three conditions simultaneously. First, it only requires a logarithmic number of qubits to represent a variable, where the logarithm is based on the number of possible values of the variable. Second, it can handle bound constraints on the variable values. Third, it can be implemented in the hard constraint model, which ensures that only feasible solutions are considered.

\subsection{Initial State}

The \textbf{initial state ($\left |ini\right \rangle$)} is a quantum state that usually represents one or more feasible solutions. The circuit for the initial state is recommanded as a constant-depth circuit, often composed of several X gates \cite{hadfield_quantum_2019}. In some literatures \cite{egger_warm-starting_2021, tate_bridging_2023, van_dam_quantum_2021}, a warm-starting method is used, which finds an initial continuous solution that is closer to the discrete optimal solution by using a classical method, thereby reducing the optimization time in subsequent steps.

\subsection{Phase-Separation Operator}
The \textbf{phase-separation operator ($U_C$)} is an operator to relate the objective function to the quantum circuit, which is usually a diagonal unitary matrix in the computational basis. The phase-separation operator can be written as $U_C = e^{-i\gamma_i C}$, where $C$ is the phase-separation Hamiltonian, and $\gamma_i$ is a circuit parameter. The encoding method and the cost function collectively determine how to construct $C$.

The general procedure to construct the phase-separation Hamiltonian $C$ involves the following steps:
\begin{itemize}
\item Reformulate the objective function to the binary objective function whose variables are binary variables $\{ \mathbf{b}_i \}$ according to the chosen encoding method.
\item Substitute $\mathbf{b}_i = \frac{1-\mathbf{z}_i}{2}$ in the equation, transforming it into a cost function with the variable vector $\mathbf{z}$.
\item Replace $\mathbf{z}$ with the Pauli Z matrix and the products between $\mathbf{z}$ in the cost function with tensor products. The phase-separation Hamiltonian $C$ is the value of the new cost function, which is a  $2^{n_z} \times 2^{n_z}$ matrix, where $n_z$ is the number of $\mathbf{z}$ or qubits.
\end{itemize}

In soft constraint models, the objective function becomes more complex, and constraints are usually added to the objective function as penalty terms. If constraints are equality constraints, the penalty term is the sum of the squares of the difference between the left and right sides of the equations. If the constraint is an inequality constraint, slack variables need to be introduced, which makes the model even more complex \cite{herman_constrained_2023, certo_comparing_2022, fernandez-lorenzo_hybrid_2021}.

In theory, multiplying the Hamiltonian $C$ by a constant $\eta$ does not change the solution to the problem. However, in real experiments, we need to use classical optimizers to optimize the phase-separation operator parameter $\gamma$ and the mixing operator $\beta$. To ensure that the optimizer's learning rate is consistent, we need to multiply a suitable parameter $\eta$ on the Hamiltonian $C$ to synchronize the steps of $\gamma$ and $\beta$. Some references have given the calculation methods for $\eta$ \cite{brandhofer_benchmarking_2022, sureshbabu_parameter_2024}.

\subsection{Mixing Operator}
The \textbf{mixing operator ($U_B$)}, also known as \textbf{mixer} is an operator that modifies the quantum state of the system and allows it to explore different regions of the solution space. The choice of the mixing operator depends on the problem structure and the optimization goal. If it is not necessary to distinguish whether the explored region is feasible or not, a common choice is the RX gate \cite{farhi_quantum_2014}. However, other operators may be more suitable for some problems, such as the XY-mixer, to transfer the quantum states among one-hot states or Dicke states \cite{hadfield_quantum_2019}. The design of the mixing operator will determine whether the QAOA algorithm is implemented in a soft constraint or a hard constraint manner. 

In hard constraint model, let $\mathcal{F}$ be the set of quantum states that represent all feasible solutions. Then, $U_B$ must satisfy the condition that there exists a parameter $\beta_i$ such that $\left \langle y \right | U_B(\beta_i) \left | z \right \rangle > 0$ for all $y, z \in \mathcal{F}$ \cite{hadfield_quantum_2019}. A common choice for $U_B$ is to use the exponential of a mixing Hamiltonian $B$, such that $U_B=e^{-i\beta_i B}$. Fuchs et al. developed a general method to find Hamiltonian $B$ given $\mathcal{F}$ \cite{fuchs_constraint_2022}. However, this method requires knowing the entire set $\mathcal{F}$, which is often impractical or impossible. Thus, this method would negate the quantum speedup. Moreover, the method of Fuchs et al. \cite{fuchs_constraint_2022} only guarantees the correct evolution of states within $\mathcal{F}$. For states outside $\mathcal{F}$, the resulting states are unpredictable. Therefore, this method cannot handle partial qubits.

However, constructing Hamiltonian $B$ is not the essential way to build a mixing operator circuit. B$\ddot{a}$rtschi et al. presented the Grover mixer \cite{bartschi_grover_2020}, which is based on the quantum gate that can transform $\left |0\right \rangle$ into an equal superposition of all feasible solutions. However, for Eq. \ref{e2}, it is still very difficult to create such a quantum gate. In this paper, we propose a novel mixing operator for QB-QAOA and use the idea of quantum chemistry circuit design \cite{whitfield_simulation_2011, yordanov_efficient_2020} to decompose it into CNOT gates and multi-control RX gates.

\subsection{Objective Function Estimation}
According to the above four components, the QAOA circuit consists of an initial circuit and $p$ layers of phase separation operator circuits and mixing operator circuits, where $p$ is the depth parameter. The final state produced by the circuit is written as $\left | \vec{\beta}, \vec{\gamma}\right \rangle =U_B(\beta _p) U_C(\gamma _p) \cdots U_B(\beta _1) U_C(\gamma _1) \left | ini\right \rangle$. Then, a classical optimizer is used to find the optimal values of $\vec{\beta}$ and $\vec{\gamma}$ that minimize the QAOA objective function $\left \langle \vec{\beta}, \vec{\gamma} \right | C \left | \vec{\beta}, \vec{\gamma} \right \rangle$. 

In some special objective functions, such as the max-cut problem and the Sherrington-Kirkpatrick Model, the objective function has an analytical solution when $p$ is small \cite{farhi_quantum_2014, farhi_quantum_2022}, so the optimal parameters can be quickly obtained by differential-based methods. However, in some general cases, it is difficult to calculate $\left | \vec{\beta}, \vec{\gamma}\right \rangle =U_B(\beta _p) U_C(\gamma _p) \cdots U_B(\beta _1) U_C(\gamma _1) \left | ini\right \rangle$. In this paper, we will present two methods for estimating it: Normal-QAOA and CVaR-QAOA \cite{barkoutsos_improving_2020}.

Normal-QAOA uses the following for estimation:

\begin{equation}
  \label{e12}
 \frac{1}{K} \sum_{k=1}^{K} \text{Cost}(\mathbf{y}^{(k)}),
\end{equation}

where $K \in \mathbb{N}$ denotes the number of measurements, which is set to $10^6$ in our simulation, $\mathbf{y}^{(k)}$ represents the value of the variable obtained by the $k$-th measurement results, and \text{Cost} is the cost function defined in Equation \ref{e4}. Therefore, Equation \ref{e12} means the average cost of all samples.

CVaR-QAOA \cite{barkoutsos_improving_2020}  uses the following for estimation:
\begin{equation}
  \label{e13}
 \frac{1}{\lceil \tau K \rceil} \sum_{k=1}^{\lceil \tau K \rceil} \text{Cost}(\mathbf{y}^{(k)}),
\end{equation}
where $\tau$ is the tail rate, and the value is 0.05 in our simulation. Equation \ref{e13} represents the average cost for the best 5\% samples. According to the Ref. \cite{barkoutsos_improving_2020}, when the optimization objective focuses on the optimal solution rather than the average performance of all solutions, CVaR-QAOA performs better than Normal-QAOA.

\subsection{Parameter Scheduling}
\label{Scheduling}

Since using classical optimization algorithms \cite{powell_direct_1994, powell_direct_1998, powell_bobyqa_2009, sack_quantum_2021, streif_training_2020} to find the global optimal parameters of $\left | \vec{\beta}, \vec{\gamma}\right \rangle =U_B(\beta _p) U_C(\gamma _p) \cdots U_B(\beta _1) U_C(\gamma _1) \left | ini\right \rangle$ is computationally expensive \cite{medvidovic_classical_2021, shaydulin_exploiting_2021, shaydulin_evaluating_2019, bittel_training_2021}, parameter scheduling methods are usually used to optimize the parameters \cite{shaydulin_multistart_2019, lotshaw_empirical_2021, galda_transferability_2021, sureshbabu_parameter_2024, shaydulin_parameter_2023}. Parameter scheduling methods usually assume that the optimal parameters satisfy certain rules, or that the optimal parameters of one instance can be inferred from the optimal parameters of instances similar to it. Parameter scheduling methods may not find the global optimal parameters, but they can find good suboptimal parameters within an acceptable time. In this paper, we will use four parameter scheduling methods in simulation experiments to estimate the optimal parameters.

\begin{itemize}
  \item \textbf{Sample20}: We randomly select 20 sets of initial parameters for training from $\gamma_i \in [-10 \pi, 10 \pi], \beta_i \in [- \pi, \pi], \forall i \in \{1, 2, \cdots p\}$, where $p$ is the number of QAOA layers. After 1000 iteration COBYLA \cite{powell_direct_1994, powell_direct_1998} for each set, we choose the set of parameters with the lowest objective function value as the optimal parameters.
  \item  \textbf{Optimized linear schedule (OLS)}: According to the theory in Ref. \cite{shaydulin_classical_2021}, the optimal parameters increase linearly with $p$. We then use the two parameters, $m_1$ and $m_2$, shown in Equation \ref{e10} to represent the original $2p$ parameters. Based on Ref. \cite{brandhofer_benchmarking_2022}, these two new parameters are optimized first. Then, according to the optimization results, we solve for the values of the original $2p$ parameters and use them as initial values for further optimization.
  \begin{equation}
    \label{e10}
    \begin{aligned}
      & \vec{\gamma}^{(\operatorname{lin}, p)}\left(m_1\right)=m_1\left(d_1^{(p)}, d_2^{(p)}, \ldots, d_p^{(p)}\right) \\
      & \vec{\beta}^{(\operatorname{lin}, p)}\left(m_2\right)=m_2\left(1-d_1^{(p)}, 1-d_2^{(p)}, \ldots, 1-d_p^{(p)}\right)\\
      &d_i^{(p)} = \frac{2i-1}{2p}\\
      \end{aligned}
  \end{equation}
  \item  \textbf{Iterative optimized linear schedule (IOLS) \cite{brandhofer_benchmarking_2022}}: The main idea of IOLS is to use the optimal parameters of a lower-layer QAOA as a guide for finding the optimal parameters of a higher-layer QAOA. Let $d_i^{(p)}$ be the coordinate of the $i$-th parameter in the $p$-layer QAOA, as defined in Equation \ref{e10}. Given the optimal parameters of the $(p-1)$-layer QAOA, $\vec{\gamma}^{(p-1)}$ and  $\vec{\beta}^{(p-1)}$, we can find two integers, $j$ and $j+1$, such that $d_j^{(p-1)}$ and $d_{j+1}^{(p-1)}$ are the nearest neighbors of $d_i^{(p)}$ among all the coordinates in the $(p-1)$-layer QAOA. IOLS assumes that the optimal parameters of the $p$-layer QAOA lie on the same line as the optimal parameters of the $(p-1)$-layer QAOA at the corresponding positions. That is, the points ($d_j^{(p-1)}$, $\gamma_j^{(p-1)}$), ($d_{j+1}^{(p-1)}$, $\gamma_{j+1}^{(p-1)}$) and ($d_{i}^{(p)}$, $\gamma_{i}^{(p)}$) are on a straight line, and so are the points ($d_j^{(p-1)}$, $\beta_j^{(p-1)}$), ($d_{j+1}^{(p-1)}$, $\beta_{j+1}^{(p-1)}$) and ($d_{i}^{(p)}$, $\beta_{i}^{(p)}$). Based on this assumption, we can use linear interpolation to estimate $\vec{\gamma}^{(p)}$ and $\vec{\beta}^{(p)}$ from $\vec{\gamma}^{(p-1)}$ and $\vec{\beta}^{(p-1)}$. These estimates are then used as initial values for further optimization.
  \item  \textbf{Iterative QAOA (IQAOA) \cite{brandhofer_benchmarking_2022}}: IQAOA is also based on the iterative method. The main idea is to use the optimal parameters of the $(p-1)$-layer QAOA as the initial values of the $p$-layer QAOA, except for the last parameter of $\vec{\gamma}^{(p)}$ and  $\vec{\beta}^{(p)}$, which is set to zero. Then, the next step is to refine the initial value with a further optimization.
\end{itemize}

\section{Quasi-Binary Based QAOA}
\label{sec4}
To solve Equation \ref{e1}, we introduce a new variable $\mathbf{y}_i= \mathbf{x}_i - L_i$ for all $i \in \{1, 2, \cdots, n\}$, and then use $\mathbf{y}_i$ to rewrite Equation \ref{e1} as follows:
\begin{equation}
  \label{e4}
  \begin{aligned}
  &\min_{\mathbf{y}}\  \sum_{i=1}^n \sum_{j=1}^n \mathbf{\sigma}_{i,j} (\mathbf{y}_i+L_i) (\mathbf{y}_j+L_j) + \sum_{i=1}^n \mathbf{\mu}_i (\mathbf{y}_i+L_i)\\
  &\textbf{s.t.} \sum_{i=1}^n \mathbf{y}_{i} =D-\sum_{i=1}^n L_i, \ 0 \leq \mathbf{y}_i \leq U_i-L_i, \   \ \mathbf{y}_i \in \mathbb{Z} \ ,\forall i \in \{1, 2, \cdots, n\}
  \end{aligned}
\end{equation}

The objective function in Equation \ref{e4} is still a quadratic form, and the discrete constraint, the bound constraint and the sum constraint have not changed significantly. After introducing the following parameters: 

\begin{equation}
  \label{e5}
  \begin{aligned}
  & \hat{D} = D-\sum_{i=1}^n L_i \\
  & R_i = U_i-L_i, \forall i \in  \{1, 2, \cdots, n\}\\
  & \mathbf{\hat{\mu}_i} = \mathbf{\mu_i}+2\sum_{j=1}^n \sigma_{i,j} L_j, \forall i \in  \{1, 2, \cdots, n\},
  \end{aligned}
\end{equation}

Equation \ref{e4} is transferred into:
\begin{equation}
  \label{e6}
  \begin{aligned}
  &\min_{\mathbf{y}}\  \sum_{i=1}^n \sum_{j=1}^n \mathbf{\sigma}_{i,j} \mathbf{y}_i \mathbf{y}_j + \sum_{i=1}^n \mathbf{\hat{\mu}}_i \mathbf{y}_i\\
  &\textbf{s.t.} \sum_{i=1}^n \mathbf{y}_{i} =\hat{D}, \ 0 \leq \mathbf{y}_i \leq R_i, \   \ \mathbf{y}_i \in \mathbb{Z} \ ,\forall i \in \{1, 2, \cdots, n\}
  \end{aligned}
\end{equation}

In this section, we will present QB-QAOA to solve the optimization model as Equation \ref{e6}.

\subsection{Quasi-Binary Encoding}

Based on Equation \ref{e6}, the encoding method should encode $\mathbf{y}_i$, within a range of 0 to $R_i$. The tradtional binary encoding is not a good choice, when $R_i \neq 2^m-1$ for $m \in \mathbb{Z}$. To address this issue, we propose the quasi-binary encoding. The quasi-binary encoding is denoted as 
\begin{equation}
  \label{e7}
  \begin{aligned}
  &\mathbf{y}_i = \sum_{j = 1}^{n_i} \sum_{k = 1}^{l_{i,j}} 2^{j-1} b_{i,j,k}, \\
  \text{where} \ \ & n_i = \lfloor \log_2 (R_i+1) \rfloor, \\
  & l_{i,j} = bin_j (R_i - 2^{n_i} + 1)+1, \\
  & b_{i,j,k} \in \{0, 1\}.
  \end{aligned}
\end{equation}

In Equation \ref{e7}, $\lfloor \log_2 (R_i+1) \rfloor$ denotes the floor of $\log_2 (R_i+1)$, and $bin_j(R_i - 2^{n_i} + 1)$ denotes the $j$-th bit of the $n_i$-length binary number of $R_i - 2^{n_i} + 1$. For example, if $R_i = 17$, then $n_i = 4$, so $R_i - 2^{n_i} + 1 = 2$, $bin(R_i - 2^{n_i} + 1) = 0010$, and $l_{i,1} = 1, l_{i,2} = 2, l_{i,3} = 1, l_{i,4} = 1$. Therefore, the encoding function is:
\begin{equation}
  \label{e8}
  \begin{aligned}
  &\mathbf{y}_i = b_{i,1,1}+2b_{i,2,1}+2b_{i,2,2}+4b_{i,3,1}+8b_{i,4,1}.
  \end{aligned}
\end{equation}

The variable is encoded with five bits: one for 1, two for 2, one for 4, and one for 8. If all bits are 1, the value is $1 \times 1+2 \times 2+1 \times 4+1 \times 8 = 17$, which is the maximum. If all bits are 0, the value is zero, which is the minimum. It is clear that the quasi-binary encoding method can represent any positive integer from 0 to $R_i$. This is because when $\mathbf{y}_i<2^{n_i}$, we can encode $\mathbf{y}_i$ with the partial bits $\{b_{i,j,1}\}$ in a binary encoding way. When $\mathbf{y}_i>=2^{n_i}$, we set all the bits in $\{b_{i,j,k}, \forall k \neq 1\}$ to 1, and encode $R_i - 2^{n_i} + 1$ with bits $\{b_{i,j,1}\}$ in a binary encoding way.

In some special cases where the gaps between $R_i$ of some variables are very large, quasi-binary encoding needs further improvement to adapt to the design of the mixing operator we will introduce later. 
The mixing operator will operate on one $2^j$-qubit and two $2^{j-1}$-qubits for all $j \leq n_i$ \footnote{We use $m$-qubit to denote the qubit that represents $m$.}. 
Hence,  we need to ensure that the total number of $2^j$-qubits in all variables should be at least 2, except for the qubit that represents the highest power of 2 (denoted as $2^{\max_i{n_i}-1}$).  
To achieve this,  after computing $n_i$ and $l_{ij}$ as Equation \ref{e7} for each $\mathbf{y}_i$, we count how many $2^{j-1}$-qubits are in all variables for each $j$ from 1 to $\max_i{n_i}-1$.  If the number is less than 2, we split one $2^{j}$-qubit into two $2^{j-1}$-qubits in the same variable. This transformation results in a finer split of integers, and does not affect the upper bounds of the variables.

As $l_{i, j} \leq 2$, we can use at most $2n_i$ qubits to represent the $\mathbf{y}_i$ value, where $n_i \leq \log_2 (R_i+1)$. Even if some qubits are splitted to adapt the mixing operator, the total number of qubits for all variables will not exceed $(2n+1) \log_2 (max_i R_i+1)$. Therefore, the number of qubits for a variable is logarithmic in the number of its possible values.

\subsection{Initial State}
To generate the initial quantum state, we must first obtain a feasible solution $\mathbf{y}$ that satisfies the conditions $\sum _{i=1}^N \mathbf{y}_i= \hat{D}$ and $\mathbf{y}_i \in [0, R_i]$. To accomplish this, we utilize the greedy allocation method, as outlined in Algorithm \ref{algo:greedy-allocate}:
\begin{algorithm}
  \caption{GreedyAllocation}
  \label{algo:greedy-allocate}
  \begin{algorithmic}[1]
    \STATE \textbf{Input:}  Upper bound array $R$ and the sum $\hat{D}$
    \STATE \textbf{Output:}  Shares array $x$
    \STATE $N \gets$ length of array $R$
    \IF{$R_0 \geq \hat{D}$}
      \STATE $x \gets$ [$\hat{D}$]+[$0$]$\times$($N-1$)
    \ELSE
      \STATE $x \gets$ [$R_0$]+GreedyAllocation($R_{1:N-1}$, $\hat{D}-R_0$)
    \ENDIF
    \RETURN $x$
  \end{algorithmic}
\end{algorithm}

Here, $R_{1:N-1}$ denotes the subarray spanning from $R_1$ to $R_{N-1}$. Subsequently, $\mathbf{y}$ is encoded into a binary sequence with quasi-binary encoding and is applyed to a quantum circuit with X gates. The greedy allocation method chooses the largest value in the range for the first few variables, which means that the first few qubits are affected by X gates in the initial circuit.

In this section, we have shown one way to create valid solutions. Any quantum state that corresponds to a valid solution under this encoding can be used. 

\subsection{Phase Separation Operator}
After substituting Equation \ref{e7} into Equation \ref{e6}, the objective function becomes a quadratic form of $b_{i,j}$. Note that $b_{i,j}^2 = b_{i,j}$ because $b_{i,j}$ is binary. Then we construct the phase-separation Hamiltonian $C$ according to the method introduced in Section \ref{e3}.

\subsection{Mixing Operator}
\label{Sec:mixer}
The mixing operator Hamiltonian of QB-QAOA can be expressed as:
\begin{equation}
  \label{e9}
  \begin{aligned}
   B &= \sum_{1 \leq i \leq n} \sum_{\substack{1 \leq j \leq n \\ r(j) = r(i)\\ j \neq i}}  X_i X_j + Y_i Y_j \\
   &+ \sum_{1 \leq i \leq n}  \sum_{\substack{1 \leq j \leq n \\ 2r(j) = r(i)}} \sum_{\substack{1 \leq k \leq n \\ 2r(k) = r(i)\\ k \neq j}}  -X_{i}Y_{j}Y_{k}+X_{i}X_{j}X_{k}+Y_{i}X_{j}Y_{k}+Y_{i}Y_{j}X_{k},
  \end{aligned}
\end{equation}
where $X_m$, $Y_m$, $Z_m$ represent the Pauli operators acting on the $m$-th qubit and $r(m)$ represents the number represented by the $m$-th qubit.

\begin{figure}[h]
  \centering
  \includegraphics[width=0.8\textwidth]{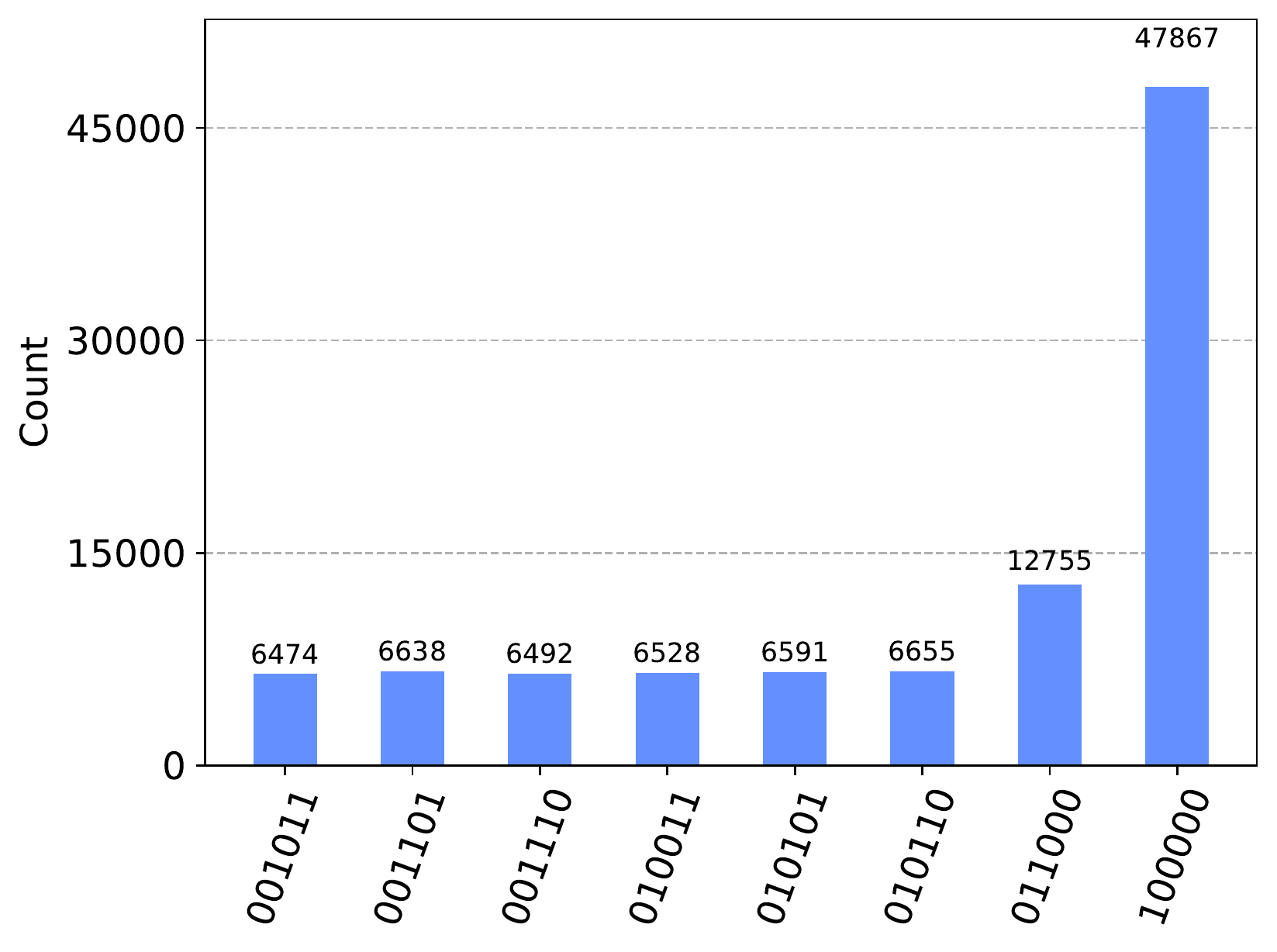}
  \caption{The measurement result of the Hamiltonian defined by Equation \ref{e9} in a 6-qubit system at t=0.3.}
  \label{fig:hist}
\end{figure}

The measurement result of the mixer $e^{-iBt}$ operated on $\left |100000 \right \rangle $ is shown in Figure \ref{fig:hist}, where Hamiltonian $B$ is defined by Equation \ref{e9} in a 6-qubit system and $t=0.3$ . In this system, the first three qubits represent 1, the fourth and fifth qubits represent 2, and the sixth qubit represents 4. Initially, the sixth qubit is initialized to $\left |1 \right \rangle$, while the others are initialized to $\left |0 \right \rangle$, which means the initial sum $\hat{D}$ is 4. After applying the Hamiltonian for $t=0.3$, the final state is measured 10,000 times. The results are shown in Figure \ref{fig:hist}, which indicates that the system only transitions to states that have a sum of 4, as expected from the sum constraint.

However, a complete mixer can not handle a large number of qubits, so we use the ring mixer to build the circuit \cite{wang_xy_2020, niroula_constrained_2022}. The circuit has two components, each corresponding to one of the terms in Equation \ref{e9}. The first component implements the ring XY mixer, and the second component implements the ring XYY mixer.

\begin{figure}[htbp]
\centering
\subfloat[Circuit realization of a two-qubit XY-mixer.\label{fig:xy-mixer}]{
\includegraphics[width=0.48 \linewidth]{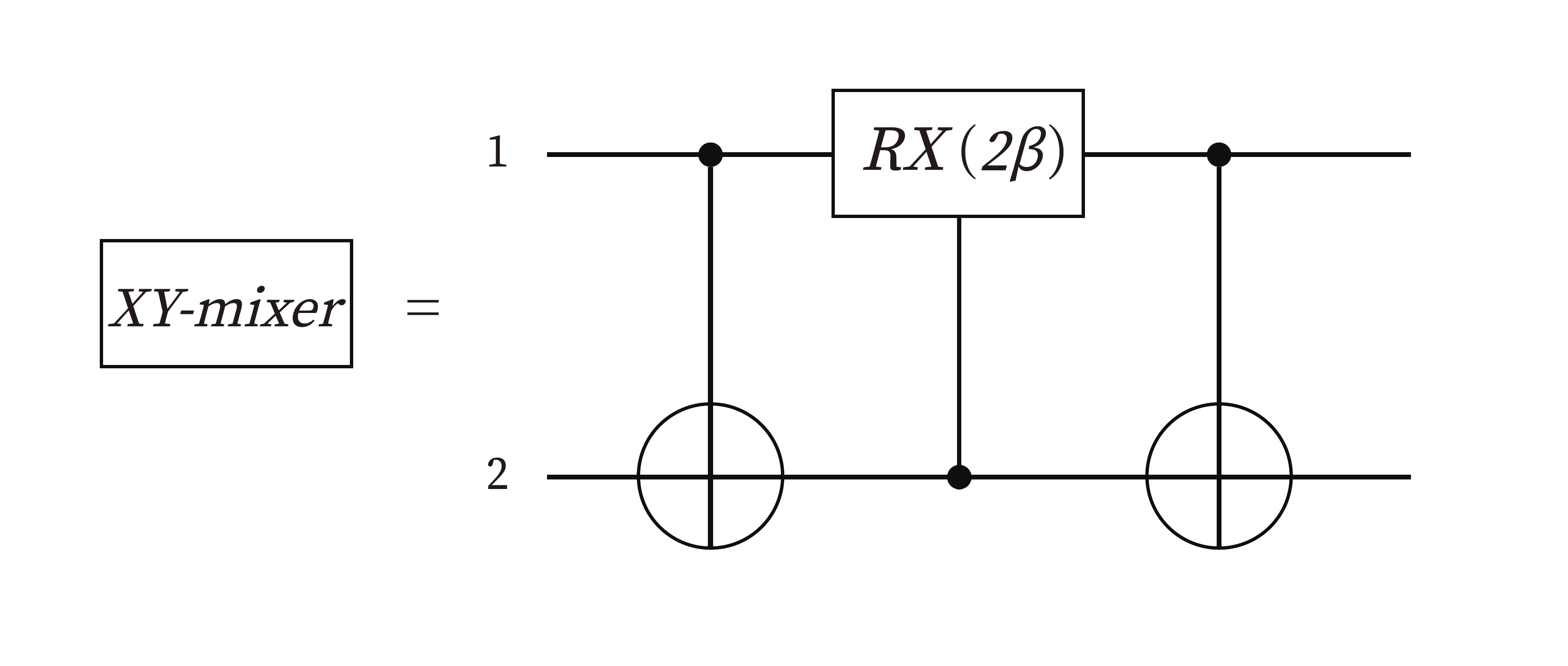}
}\hfill
\subfloat[Ring XY-mixer among 6 qubits.\label{fig:ring-xy-mixer}]{
\includegraphics[width=0.48 \linewidth]{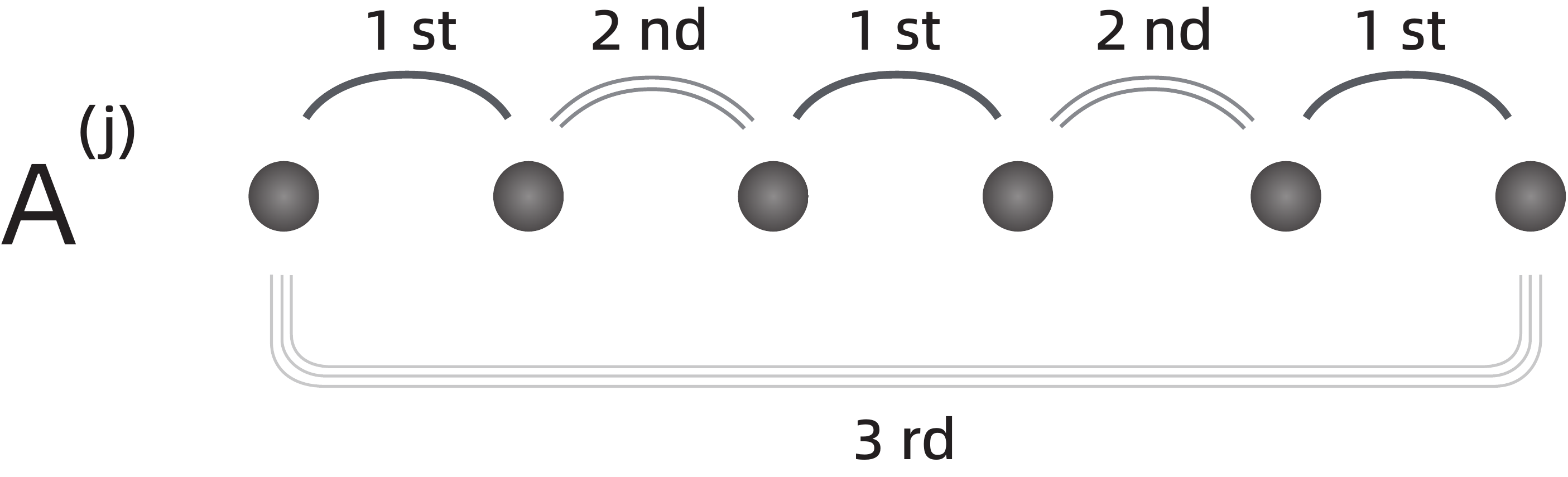}
}\hfill
\caption{The implementation of the XY-mixer.} % 总标题
\label{fig:two-xy-mixers} % 整个组图的标签
\end{figure}

Figure \ref{fig:xy-mixer} illustrates the circuit realization of a two-qubit XY-mixer, which partially swaps the states $\left |01 \right \rangle$ and  $\left |10 \right \rangle$. The XY-mixers operate on qubits that have the same integer value in the quasi-binary encoding. For example, if we have $A^{(j)}$ as the set of qubits to represent $2^j$, the ring XY-mixer will be operated on these qubits in a certain order based on the literature \cite{hadfield_quantum_2019}. Figure \ref{fig:ring-xy-mixer} explains how this works when $A^{(j)}$ has 6 qubits. The implementation of the ring XY-mixer is divided into three rounds, and in each round, the XY-mixers act on the qubits at the ends of the arc in Figure \ref{fig:ring-xy-mixer}.

\begin{figure}[htbp]
  \centering
  \vspace{0.5cm}
  \subfloat[Circuit realization of a three-qubit XYY-mixer.\label{fig:xyy-mixer}]{
  \includegraphics[width=0.48 \linewidth]{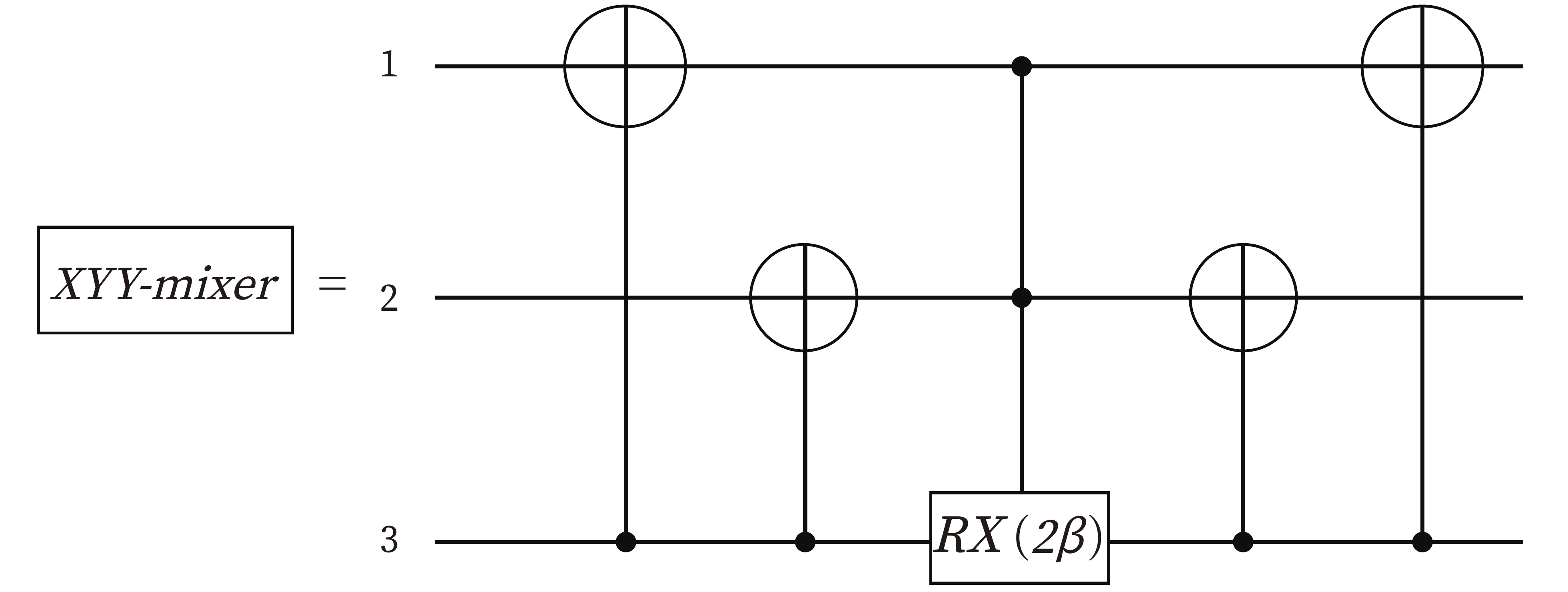}
  }\hfill
  \subfloat[Ring XYY-mixer among 12 qubits.\label{fig:ring-xyy-mixer}]{
  \includegraphics[width=0.48 \linewidth]{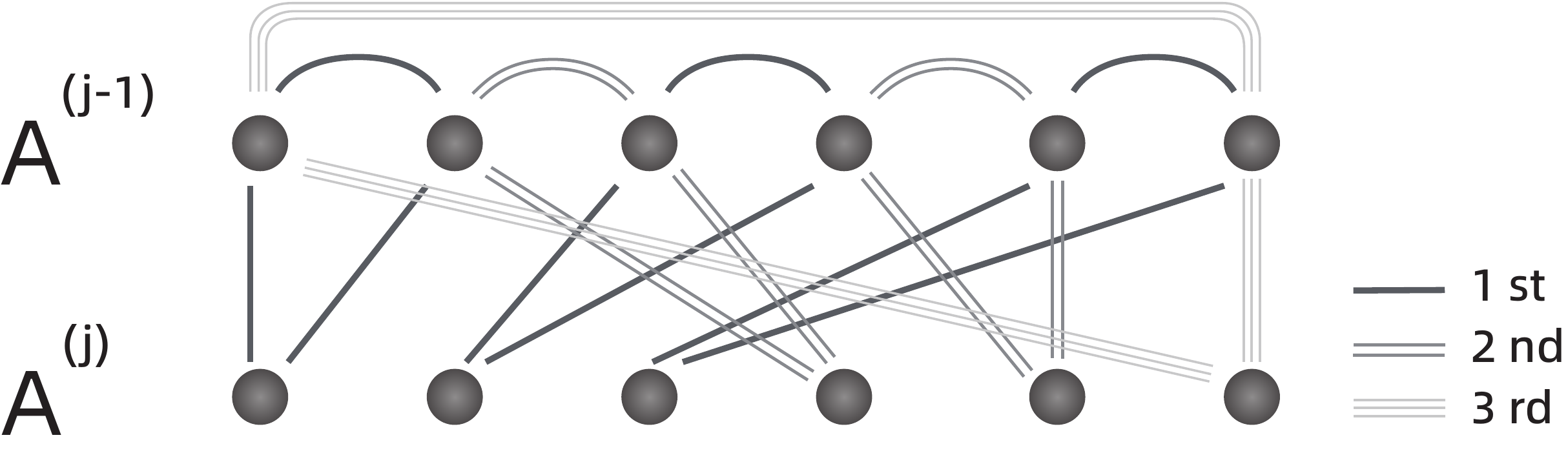}
  }\hfill
  \caption{The implementation of the XYY-mixer.} % 总标题
\end{figure}

Figure \ref{fig:xyy-mixer} shows how to build a three-qubit XYY-mixer circuit, which partially swaps the states $\left |011 \right \rangle$ and  $\left |100\right \rangle$. XYY-mixers work on two qubits to represent $2^{j-1}$ (the first and second qubits) and one qubit to represent $2^j$ (the third qubit). Given that the XYY-mixers work on the qubits in $A^{(j)}$ and $A^{(j-1)}$, which are the sets of qubits that represent $2^j$ and $2^{j-1}$, respectively. Figure \ref{fig:ring-xyy-mixer} shows the order of applying the XYY-mixer to these qubits. The ring XYY-mixer has also three rounds, like the ring XY-mixer. The way of choosing two qubits from $A^{(j-1)}$ is the same as the XY-mixer in Figure \ref{fig:ring-xy-mixer}. After picking the two qubits from $A^{(j-1)}$, the XYY-mixer picks one single qubit from $A^{(j)}$ that have not been picked yet.

We assume that both $A^{(j-1)}$ and $A^{(j)}$ have six qubits in Figure \ref{fig:ring-xyy-mixer}. But in reality, the number of qubits in $A^{(j)}$ does not depend on the number of qubits in $A^{(j-1)}$, so there might be more or less qubits in $A^{(j)}$ when we pair them up. If there are not enough qubits in $A^{(j)}$, we can use the qubits that have been picked before for the XYY-mixer. If there are extra qubits in $A^{(j)}$, we can ignore them.

Algorithm \ref{aaaa2} gives the construction process of the whole mixer circuit. The pairing methods and order shown in Figure \ref{fig:ring-xy-mixer} and Figure \ref{fig:ring-xyy-mixer} are also given more clearly in pseudocode. The depth of the whole mixing circuit is $\max (\lceil \frac{L_{j-1}}{L_j}\rceil,3)+3$, where $L_{j-1}$ and $L_j$ are the size of $A^{(j-1)}$ and $A^{(j)}$.

\begin{algorithm}[htbp]
  \caption{MixingCircuit}
  \label{aaaa2}
  \begin{algorithmic}[1]
    \STATE Let $J$ represent the maximum power of 2 that can be expressed by all qubits
    \FOR{$j \gets 0$ to $J$}
    \STATE Let $A^{(j)}$ represent the set of all $2^j$-qubit sets of assets
    \STATE $L \gets$ size of $A^{(j)}$
    \FOR{$i \gets 1$ to $L$}
    \IF{$i$ is odd and $i<L$}
    \STATE Apply \fbox{XY-mixer} on $i$-th qubit and $i+1$-th qubit in $A^{(j)}$
    \ENDIF
    \ENDFOR
    \FOR{$i \gets 1$ to $L$}
    \IF{$i$ is even and $i<L$}
    \STATE Apply \fbox{XY-mixer} on $i$-th qubit and $i+1$-th qubit in $A^{(j)}$
    \ENDIF
    \ENDFOR
    \STATE Apply \fbox{XY-mixer} on the last qubit and the first qubit in $A^{(j)}$
    \ENDFOR
    \FOR{$j \gets J$ to $1$ by $-1$}
    \STATE $k \gets 1$
    \STATE $L \gets$ size of $A^{(j-1)}$
    \STATE $M \gets$ size of $A^{(j)}$
    \FOR{$i \gets 1$ to $L$}
    \IF{$i$ is odd and $i<L$}
    \STATE Apply \fbox{XYY-mixer} on $i$-th qubit and $i+1$-th qubit in $A^{(j-1)}$ and $k$-th qubit in $A^{(j)}$
    \STATE $k \gets (k+1)\% M$
    \ENDIF
    \ENDFOR
    \FOR{$i \gets 1$ to $L$}
    \IF{$i$ is even and $i<L$}
    \STATE Apply \fbox{XYY-mixer} on $i$-th qubit and $i+1$-th qubit in $A^{(j-1)}$ and $k$-th qubit in $A^{(j)}$
    \STATE $k \gets (k+1)\% M$
    \ENDIF
    \ENDFOR
    \STATE Apply \fbox{XYY-mixer} on the last qubit and the first qubit in $A^{(j-1)}$ and $k$-th qubit in $A^{(j)}$
    \ENDFOR
    \end{algorithmic}
\end{algorithm}

\section{Numerical Simulation}
\label{sec5}
\subsection{Data and parameter}
\label{nasdaq}
\begin{figure}[h]
  \centering
  \includegraphics[width=0.8\textwidth]{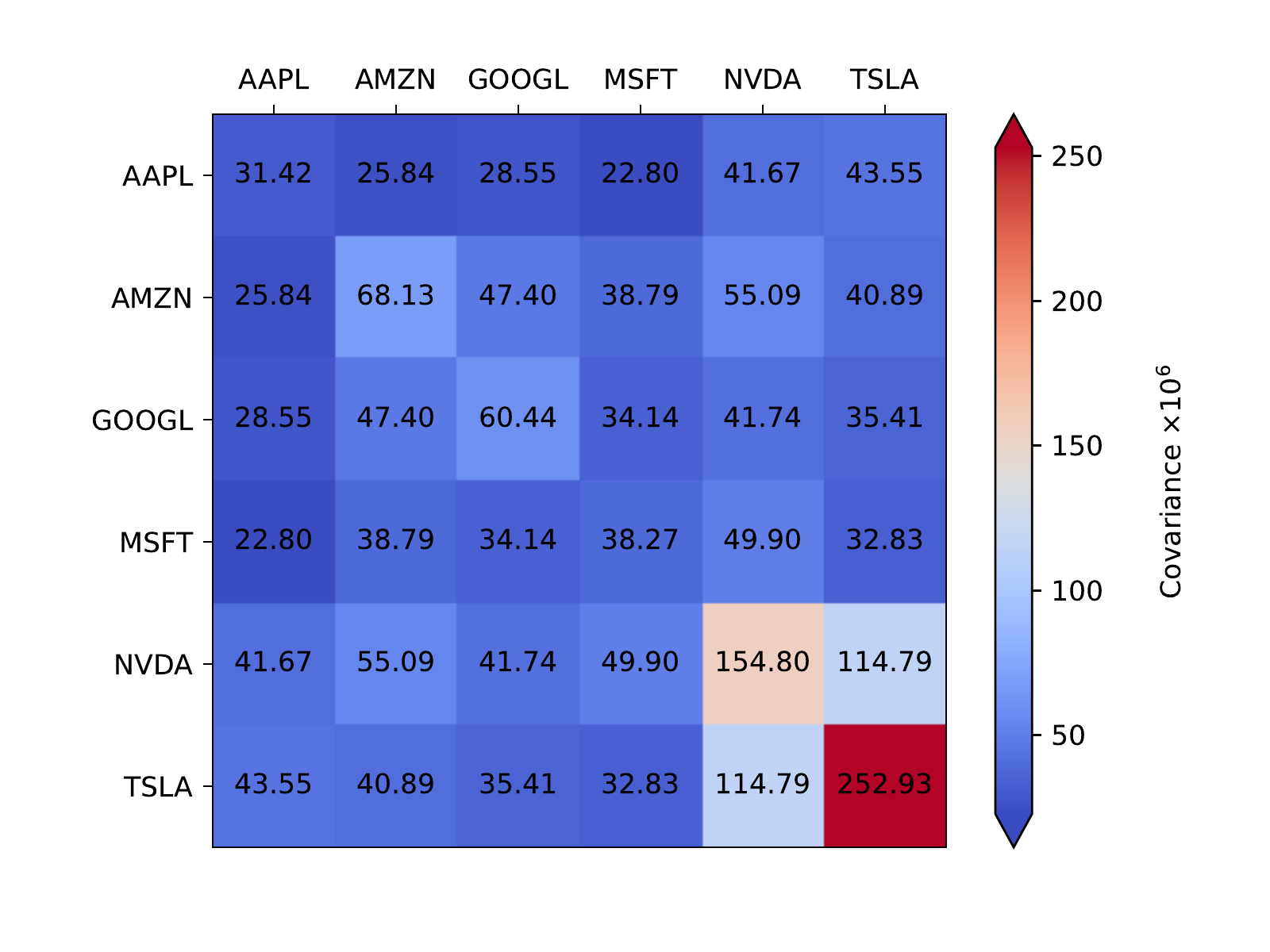}
  \caption{Daily asset returns covariance as $\sigma_{ij} \times 10^6$ for 6 stocks selected from NASDAQ.}
  \label{fig:cov}
\end{figure}

We applied the QB-QAOA method to optimize a portfolio of six NASDAQ stocks (AAPL, AMZN, GOOGL, MSFT, NVDA, TSLA) using Qiskit. We used the historical returns rates of these stocks as the input data. Table \ref{table:stock_returns} shows the expected returns and Figure \ref{fig:cov} shows the covariance matrix of the stocks.

\begin{table}[htbp]
  \centering
  \begin{tabular}{|c|cccccc|}
  \hline
  \textbf{Stock} & AAPL & AMZN & GOOGL & MSFT & NVDA & TSLA \\
  \hline
  \textbf{Returns Expectation} & 0.134  \textperthousand  & 0.354 \textperthousand& -1.582 \textperthousand& -0.152 \textperthousand& 6.261 \textperthousand& 2.187 \textperthousand\\ \hline

  \hline
  \end{tabular}
  \caption{Daily asset returns expectation for 6 stocks selected from NASDAQ.}
  \label{table:stock_returns}
\end{table}

The risk factor $q$ is setted as 18.415, so according to the Appendix \ref{a1}, we get the minimum target return $\mu$ of exactly 2 \textperthousand. The values of the precision parameter $\alpha$, the lower bound $l$, and the upper bound $u$ depend on the simulation content and may change accordingly.

\subsection{Qubits count}
\begin{figure}[htbp]
  \centering
  \includegraphics[width=0.8\textwidth]{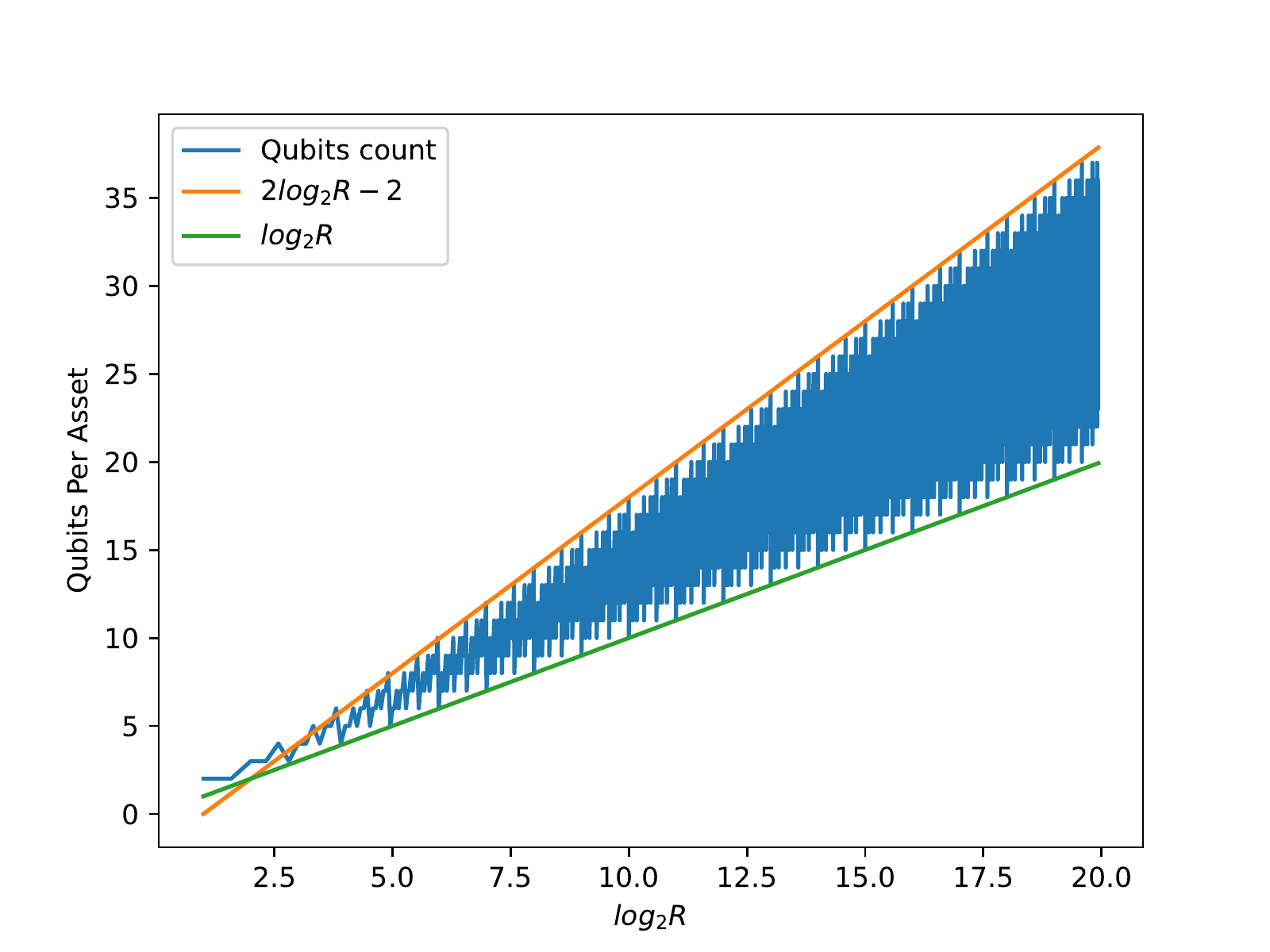}
  \caption{The relationship between the number of qubits to encode per asset and $\log_{2}(R)$.} 
  \label{fig:qubits}
  \end{figure}

To study how the range width $R$ affects the number of qubits to encode the assets shares, we assume that all six assets have the same range, from 0 to 1. We can change the range width by adjusting the precision parameter $\alpha$, where $R = \frac{u}{\alpha}- \frac{l}{\alpha} = \frac{1}{\alpha}$. Figure \ref{fig:qubits} shows that the number of qubits for each asset is between $\log_2 R$ and $2\log_2 R-2$. For example, when $R=1000$ ($\alpha=1$ \textperthousand) , we need 16 qubits for each asset. By contrast, Hodson encoding needs 1000 qubits, one-hot encoding needs 1001 qubits, the binary encoding that does not support bound constraints and the Palmer encoding that is only applicable to soft constraints use 10 qubits.

\subsection{Mixability}
\label{sec::Mixability}

\begin{figure}
  \centering
  \subfloat[$p=1$.\label{fig:mixer1}]{
  \includegraphics[width=0.48 \linewidth]{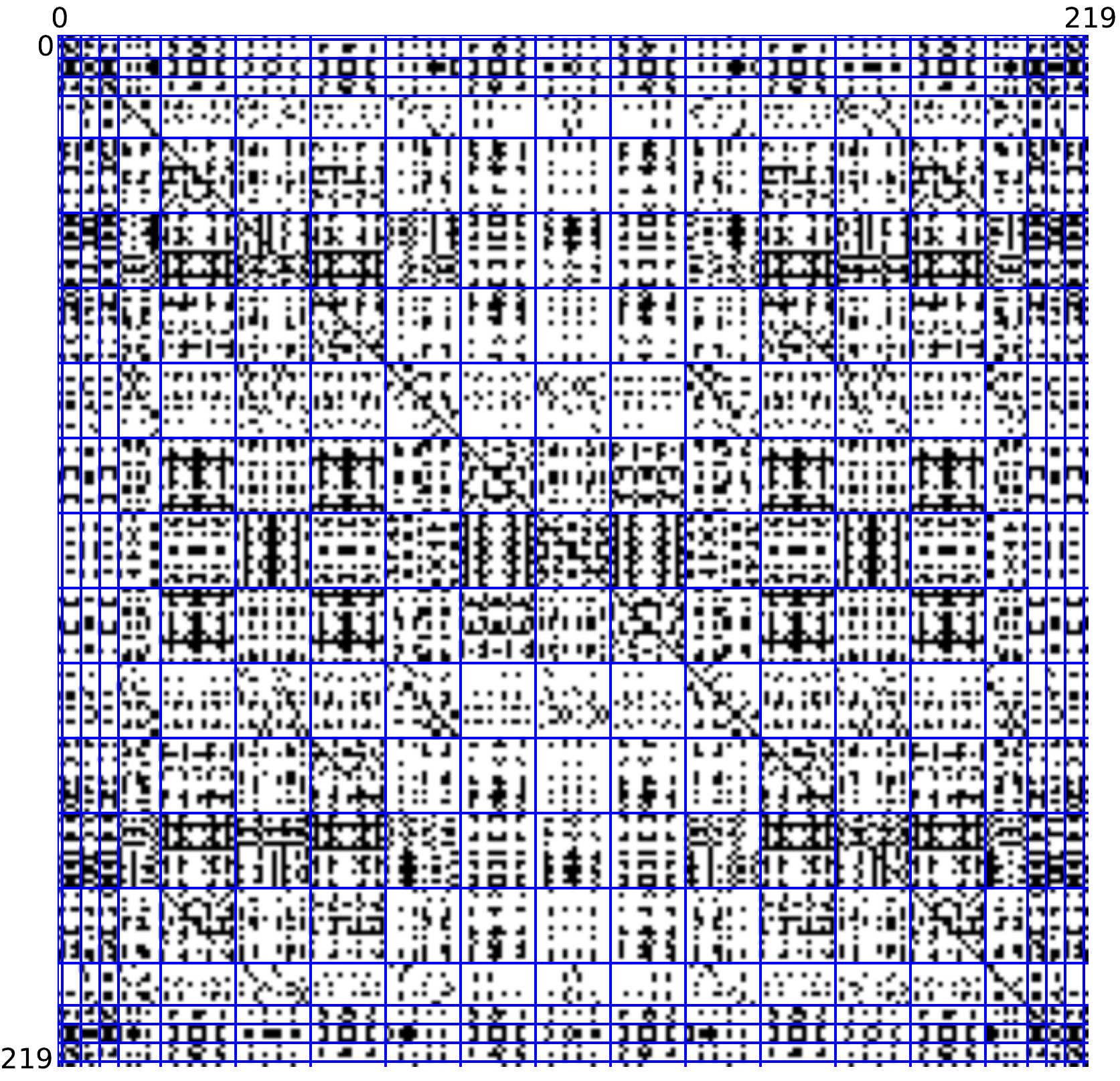}
    }\hfill
  \subfloat[$p=2$.\label{fig:mixer2}]{
  \includegraphics[width=0.48 \linewidth]{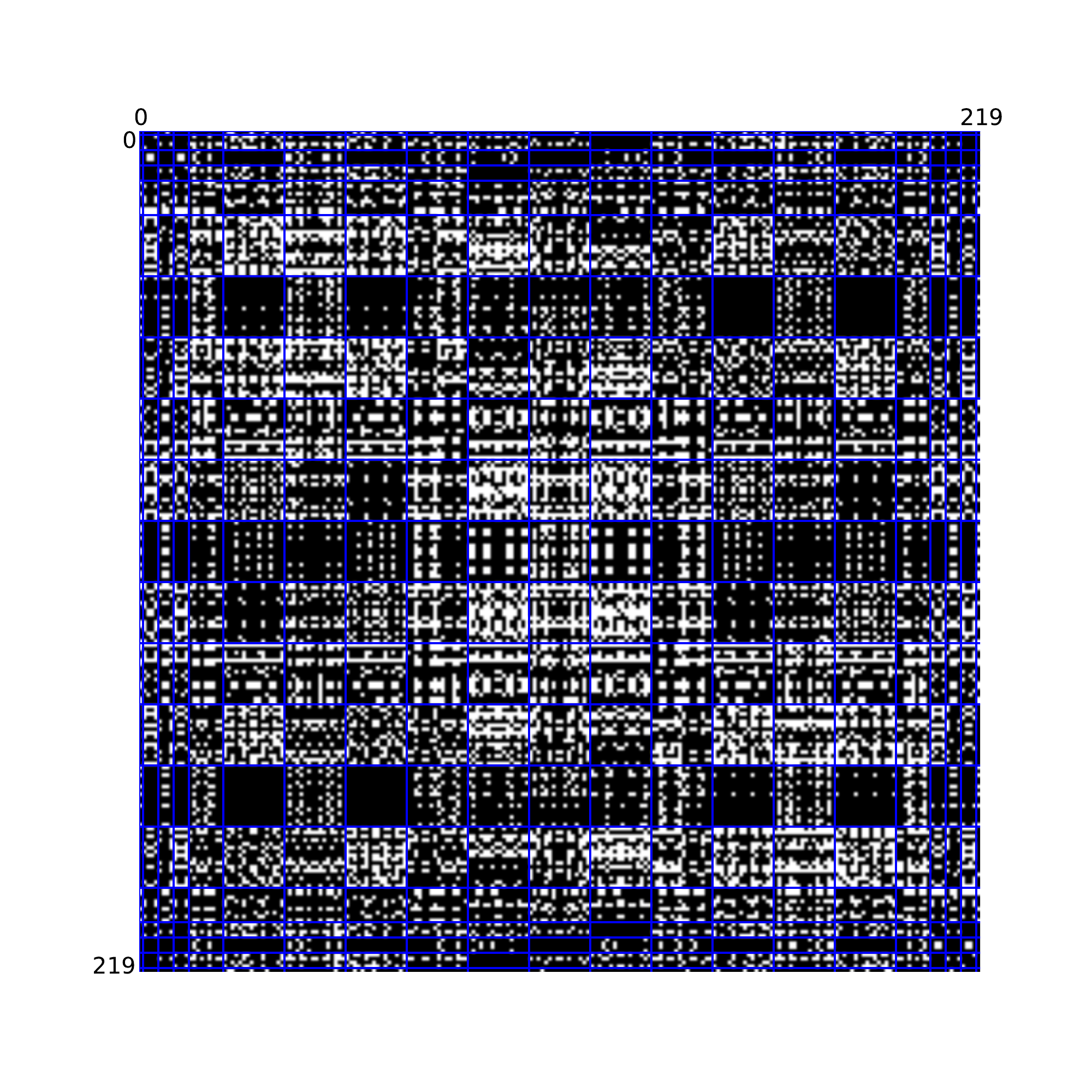}
    }\hfill
  \subfloat[$p=4$.\label{fig:mixer3}]{
  \includegraphics[width=0.48 \linewidth]{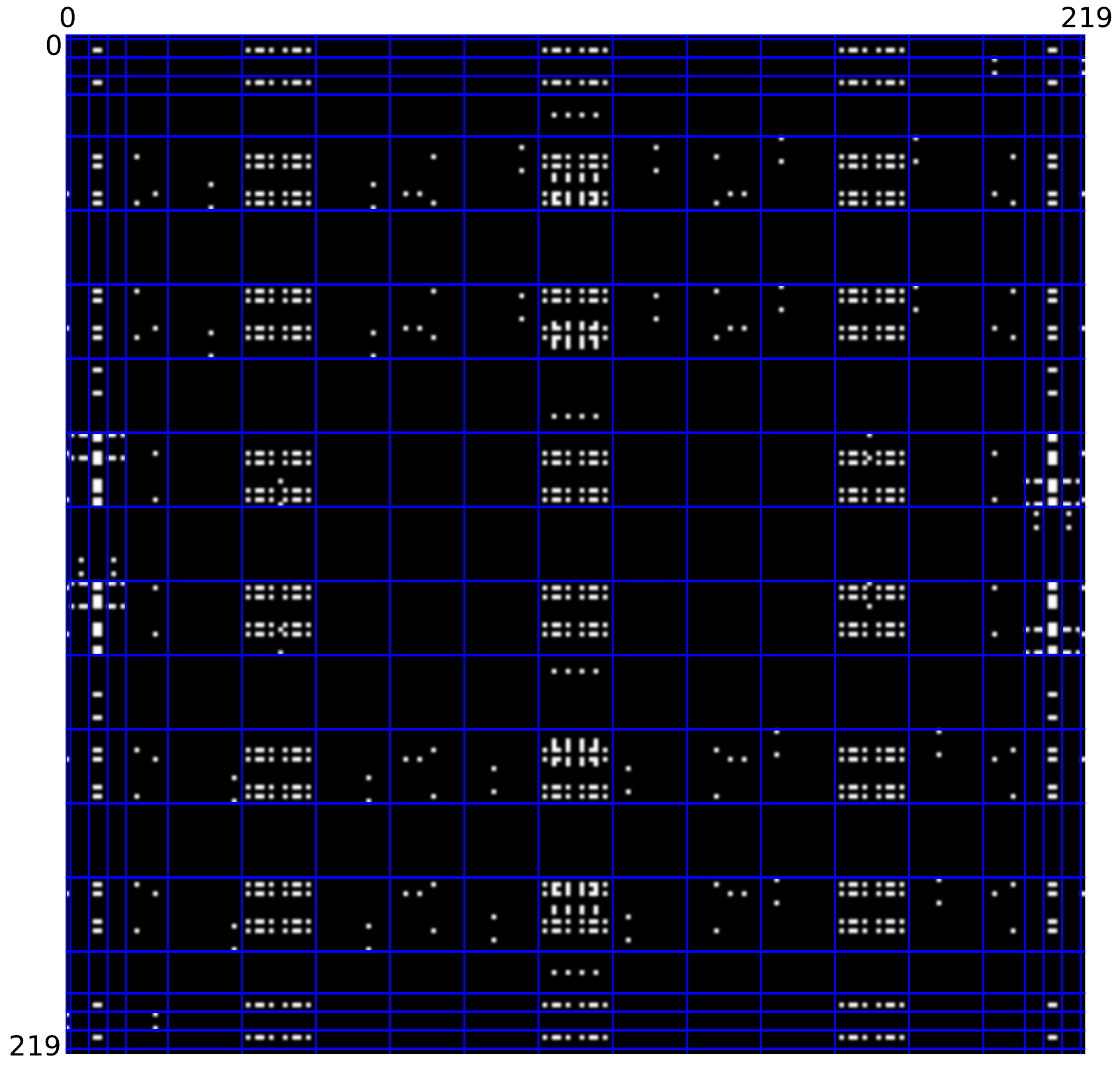}
      }\hfill
  \subfloat[$p=8$.\label{fig:mixer4}]{
  \includegraphics[width=0.48 \linewidth]{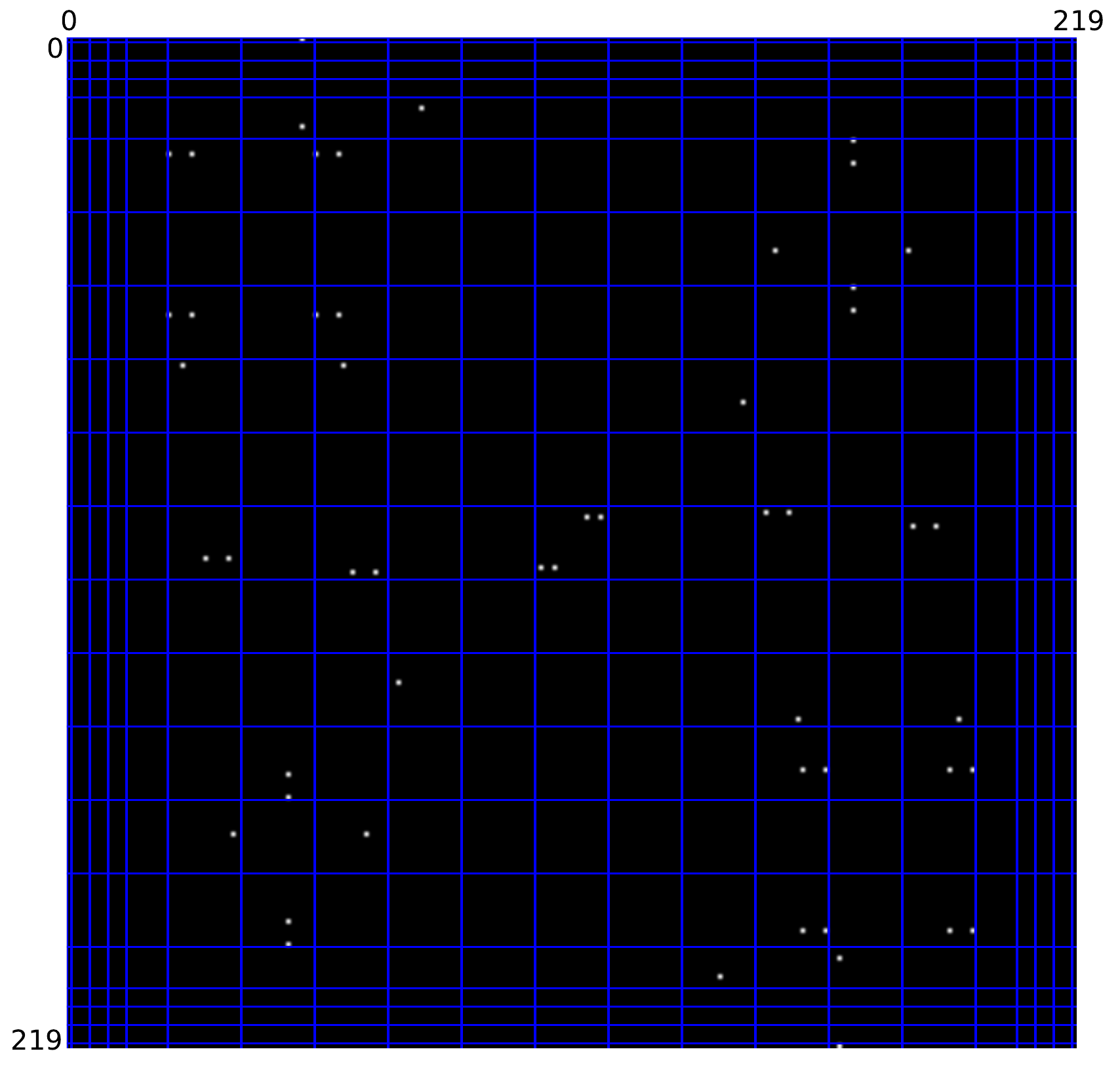}
      }\hfill
  \caption{The gray scale image of $M^p$}
  \label{fig:mixer}
\end{figure}

Mixability is defined as a property of a mixer that quantifies its ability to explore the feasible region of an optimization problem. A mixing operator should be able to move from one feasible quantum state to any other feasible quantum states, given suitable circuit parameters $\vec{\beta}$ and enough layers of mixers. To quantify the performance of a mixer, we use the following metric:

\begin{equation}
  \label{e11}
  M^p (\left | \psi \right \rangle,  \left | \phi \right \rangle ) = \max_{\vec{\beta}}  \left | \left \langle \phi \right | U_B(\beta_p) U_B(\beta_{p-1}) \cdots U_B(\beta_1) \left | \psi \right \rangle \right |,
\end{equation}

where $\left | \phi \right \rangle$ is the target quantum state representing the solution we want to explore, $\left | \psi \right \rangle$ is the initial quantum state, and $p$ is the number of layers of $U_B$. This equation represents the square root of the fidelity between the final state $U_B(\beta_p) U_B(\beta_{p-1}) \cdots U_B(\beta_1) \left | \psi \right \rangle$ and the target state $\left | \phi \right \rangle$ under an optimal sequence of parameters $\vec{\beta}$. Since $\left | \phi \right \rangle$ and $\left | \psi \right \rangle$ are quantum states on the computational basis, this equation also represents the modulus of the amplitude of $\left | \phi \right \rangle$ on the final state which is a superposition state.

In our simulation, we focus on two stocks, NVDA and TSLA, where both the upper and lower bounds are set to 1 and -1, respectively. The precision parameter $\alpha$ is fixed at 0.1, implying that $R_i=20$, where $i\in \{\text{NVDA},\text{TSLA}\}$. Each asset is represented by six qubits in the quasi-binary encoding, consisting of two 1-qubits, one 2-qubit, two 4-qubits, and one 8-qubit. The sum constraint is $\mathbf{y}_\text{NVDA}+\mathbf{y}_\text{TSLA}=20$. Accordingly, the set of feasible quantum states is expressed as $\mathcal{F} = \{\psi | f(\psi_{1:6})+f(\psi_{7:12})=20 \}$, where $f$ is the decoding function, $\psi_{1:6}$ is the binary sequence of the first six qubits and $\psi_{7:12}$ is the binary sequence of the last six qubits. There are 21 portfolios that satisfy the constraint, and there are 220 feasible encodings to encode these 21 portfolios under quasi-binary encoding. In Figure \ref{fig:mixer}, the blue line is used to distinguish different portfolios, and pixels are used to distinguish each different encodings.

Since calculating the global optimal parameters of Equation \ref{e11} is a challenging task, we use a discrete set of points as $\vec{\beta}$ to estimate the value of $M^p$. For a parameter vector of size $p$, we choose the parameter subject to $\beta_1 = \beta_2 = \cdots = \beta_p = \frac{i}{4} \pi, i =1, 2, \cdots, 7$. Then, we use the maximum value of these parameters as the estimation of Equation \ref{e11}. 

Figure \ref{fig:mixer} displays whether the estimation is greater than a preset threshold $\epsilon=0.001$, when $p$ is 1, 2, 4, and 8 respectively. When the estimation of $M^p (\left | \psi \right \rangle,  \left | \phi \right \rangle )$ is greater than $\epsilon$, the pixels of the row representing $| \phi \rangle$ and the column representing $| \psi \rangle$ are plotted as black, which means an effective conversion between $| \phi \rangle$ and $| \psi \rangle$, otherwise as white. Therefore, the entire figure consists of $220 \times 220$ pixels, corresponding to all 220 feasible encodings of the two assets. 

By comparing the grayscale map changes under different $p$ values, we can observe that the mixability of encoding pairs that exceed the threshold $\epsilon$ is increasing. In fact, if the threshold $\epsilon$ is set to 0.0001, the entire grayscale image will become black at $p=8$, which means that any two feasible quantum states will undergo effective conversion.

\subsection{Real instance}
\label{sec::realsimulation}

In the real instance, to evaluate the performance of different algorithms to find the optimal solution, we use the approximation ratio as a metric. The approximation ratio, denoted as $ar$, is defined as 

\begin{equation}
  \label{e14}
  ar = \frac{C_W-C_A}{C_W-C_O},
\end{equation}
where $C_O$ and $C_W$ represent the costs of the optimal and worst solutions, respectively. They are obtained by the Brute-Force method. $C_A$ represents the average cost of all samples. For QAOA, $C_A$ is shown as Equation \ref{e12} or Equation \ref{e13} and for the Brute-Force method, the expectation is based on that the probability of each different solution appearing is the same.

During the parameter optimization process, we use COBYLA as the classical optimizer to find the best parameter. However, COBYLA may get trapped in different local optima depending on the initial parameter values. To overcome this challenge, we apply four different parameter scheduling methods described in Section \ref{Scheduling} to determine the initial parameters.

\subsubsection{Instance 1}
\begin{figure}[htbp]
  \centering
  \includegraphics[width=0.8\textwidth]{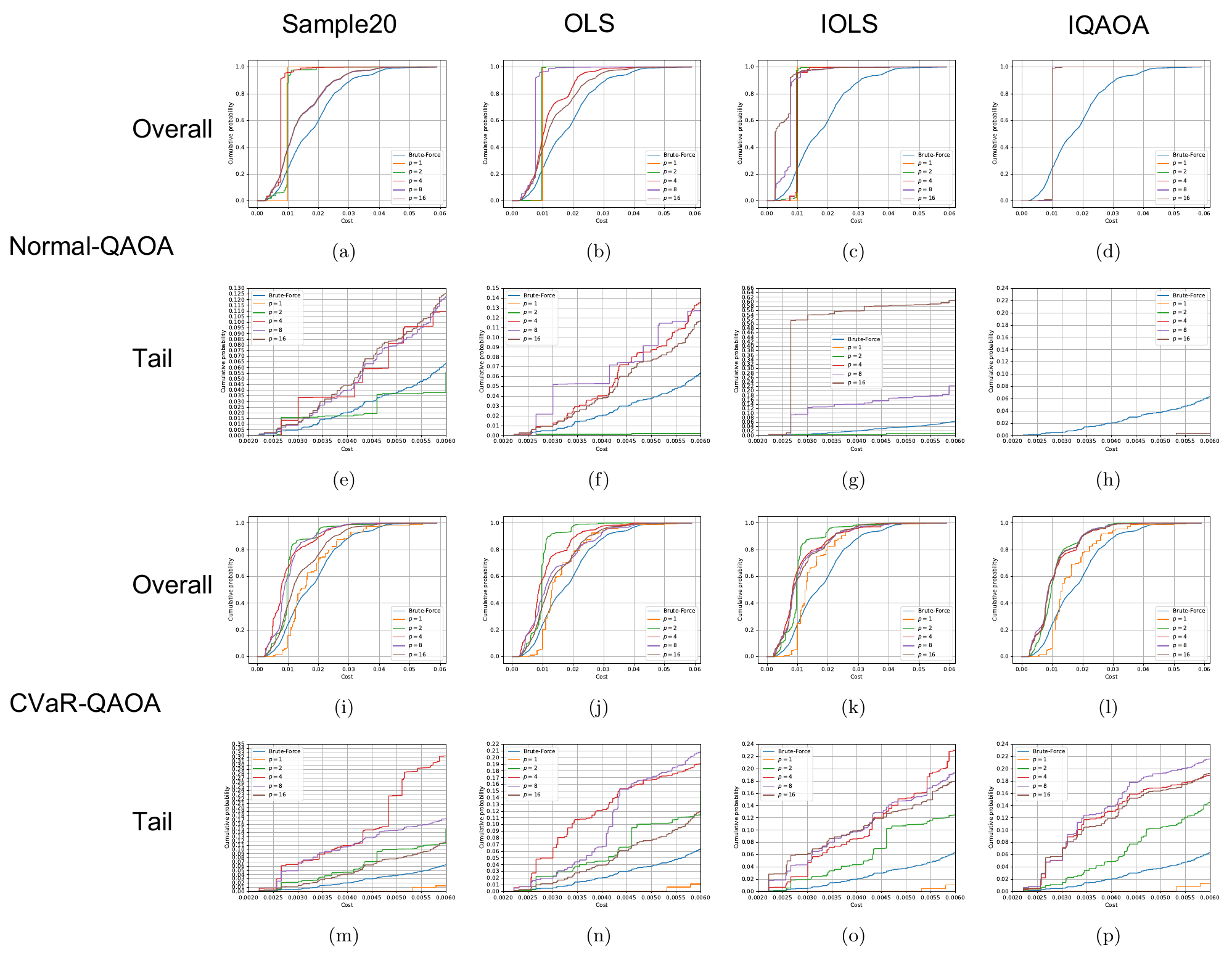}
  \caption{Cumulative distribution probability of solution costs from different types of QAOA algorithms.} 
  \label{fig:instance1}
\end{figure}

In the first case, we select six stocks shown in Section \ref{nasdaq} as the stock pool. We set $\alpha=0.5$, with upper and lower bounds of 1 and -1 for all assets, requiring a total of 18 qubits for the experiment. We also conduct different simulations for $p=1$, $p=2$, $p=4$, $p=8$, and $p=16$, respectively.

Figure \ref{fig:instance1} shows the cumulative distribution probability of the cost of each different solution under Normal-QAOA and CVaR-QAOA with four parameter estimation methods and five different depth parameters $p$. Since the portfolio optimization problem cares more about the cumulative distribution probability of the part of the solution with the lowest cost, we separately list the cumulative distribution probability of the approximately 6.4\% optimal solution as the tail.

In terms of tail performance, the CVaR-QAOA method outperforms the Normal-QAOA method significantly. The CVaR-QAOA method is superior to the Brute-force method when $p$ exceeds 2. However, under Normal-QAOA, the IOLS method needs to show its superiority at $p=8$, and the IQAOA method cannot show its superiority even at $p=16$. Comparing the IOLS and IQAOA methods under Normal QAOA, it can be found that these two parameter scheduling methods based on iterative methods are very easy to fall into local optimal solutions. As the value of $p$ increases, the distribution of the solutions of large $p$ and small $p$ are highly similar, but this problem does not exist in the CVaR-QAOA method. Therefore, under the quasi-binary encoding, it is recommended that the IOLS and IQAOA methods be used under the CVaR-QAOA framework. In addition, the performance at $p=16$ is worse than that at $p=4$ and $p=8$ in most cases. The explanation for this is that the parameter count for $p=16$ (32 parameters) is too large to find a better solution by classical COBYLA algorithm with 1000 iterations. However, under IOLS in Normal-QAOA method, the performance at $p=16$ is exceptionally good, jumping to about 52\% probability at a cost of about 0.027. This also shows that once a better parameter is found at $p=16$, the probability distribution of the solution is significantly improved.

\begin{figure}[htbp]
  \centering
  \includegraphics[width=0.9\textwidth]{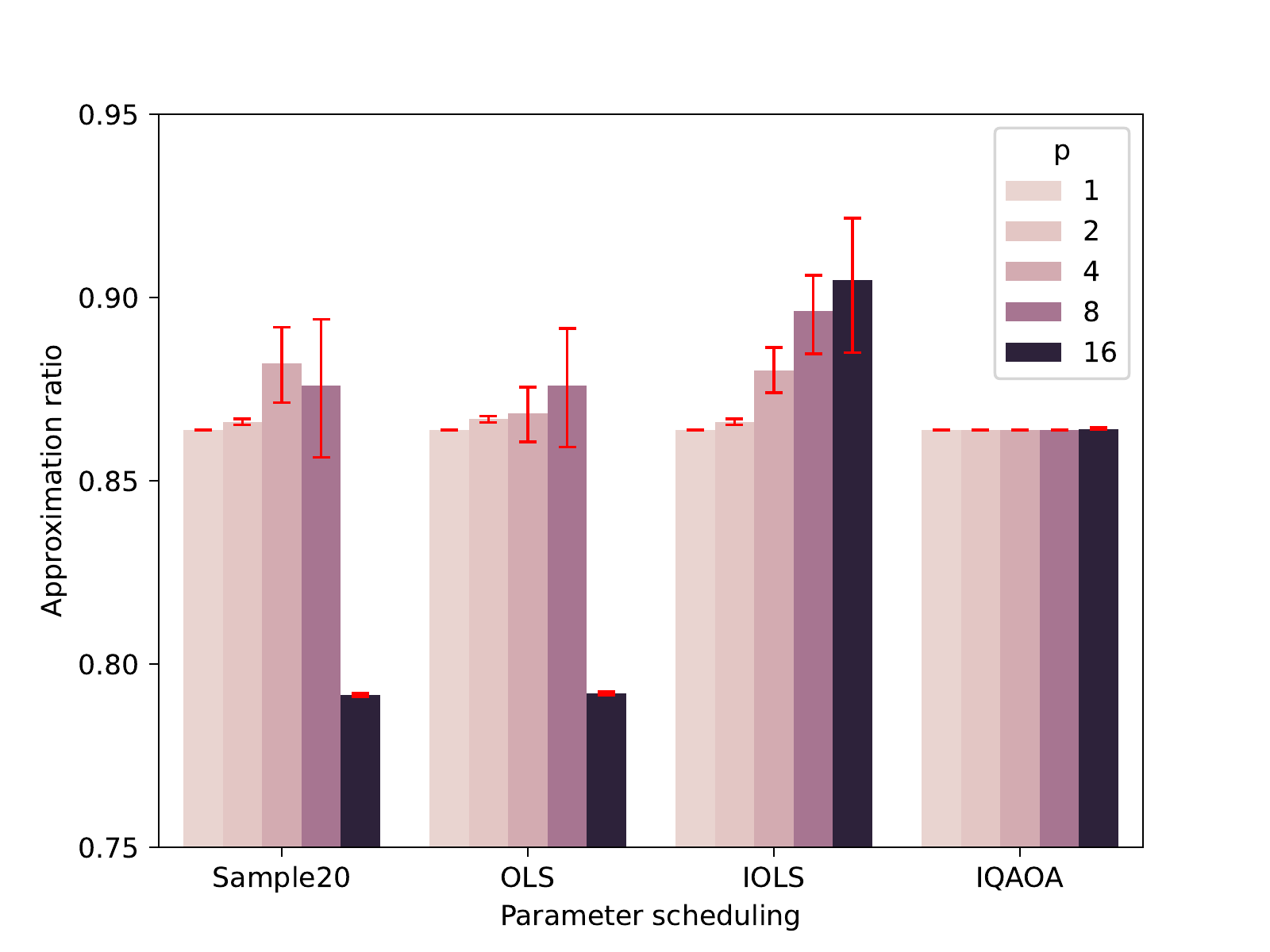}
  \caption{Approximation ratio of Normal-QAOA under different parameter scheduling methods and $p$ values.}
  \label{fig:normal_qaoa_ar}
\end{figure}

\begin{figure}[htbp]
  \centering
  \includegraphics[width=0.9\textwidth]{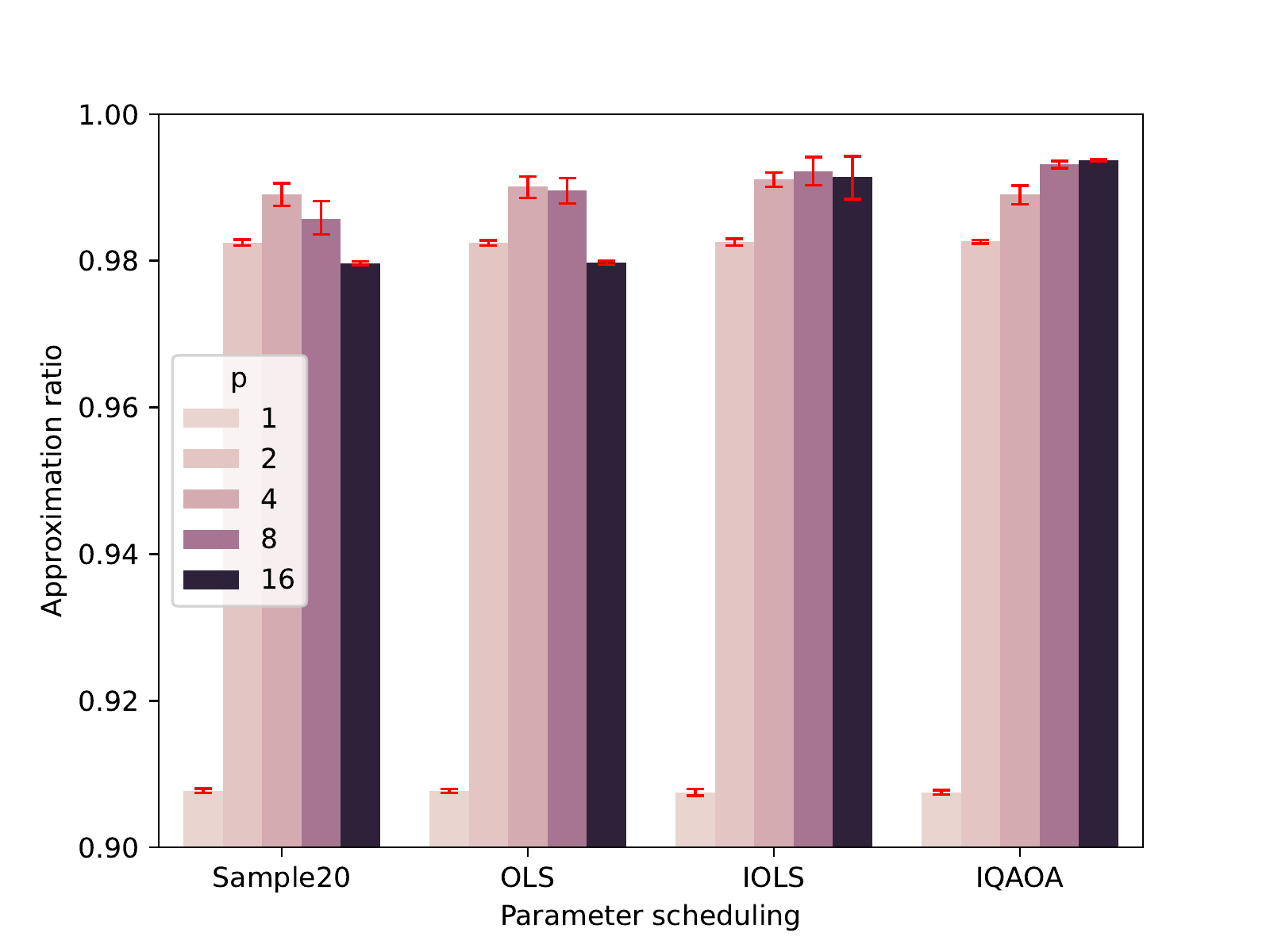}
  \caption{Approximation ratio of CVaR-QAOA under different parameter scheduling methods and $p$ values.}
  \label{fig:CVaR_qaoa_ar}
\end{figure}

To avoid errors caused by a single experiment, we repeated the above experiment 33 times. Figures \ref{fig:normal_qaoa_ar} and \ref{fig:CVaR_qaoa_ar} show the approximation ratios of Normal-QAOA and CVaR-QAOA under different parameter scheduling methods and $p$ values, with error bars representing the 95\% confidence interval. The approximation ratio of Normal-QAOA is between 0.791 and 0.905. It can also be seen that the approximation ratio decreases when $p=16$ under Sample20 and OLS parameter scheduling methods, and the error range is almost non-existent. This indicates that the experiment has encountered the barren plateau phenomenon at $p=16$, and small-scale repetitive experiments cannot overcome this phenomenon. The approximation ratio of IOLS parameter scheduling method increases with the depth of the circuit, and the confidence interval also expands, indicating that the IOLS method can resist the barren plateau phenomenon to a certain extent. The IQAOA method shows the same behavior as shown in Figure \ref{fig:instance1}, where the solution falls into a local optimum during the iterative process, and the approximation ratio remains almost unchanged from $p=1$ to $p=16$, with little fluctuation. The approximation ratio of CVaR-QAOA ranges from 0.907 to 0.994. Among the four scheduling methods, the IQAOA method, which performs the worst under Normal-QAOA, is the best schedualing method when $p=8$ and $p=16$. The explanation for this is that by using only the best 5\% samples as the optimization target, perturbations can be added during the optimization process to overcome the problem of local optima caused by the iterative method. Therefore, we believe that in the portfolio optimization scenario, the QAOA method based on quasi-binary encoding should adopt the CVaR-QAOA estimation, and the IQAOA or IOLS parameter scheduling method. The value of $p$ is recommended to be above 8, so that the approximation ratio of the solution is above 0.99.

\subsubsection{Instance 2}
\begin{figure}[htbp]
  \centering
  \includegraphics[width=0.9\textwidth]{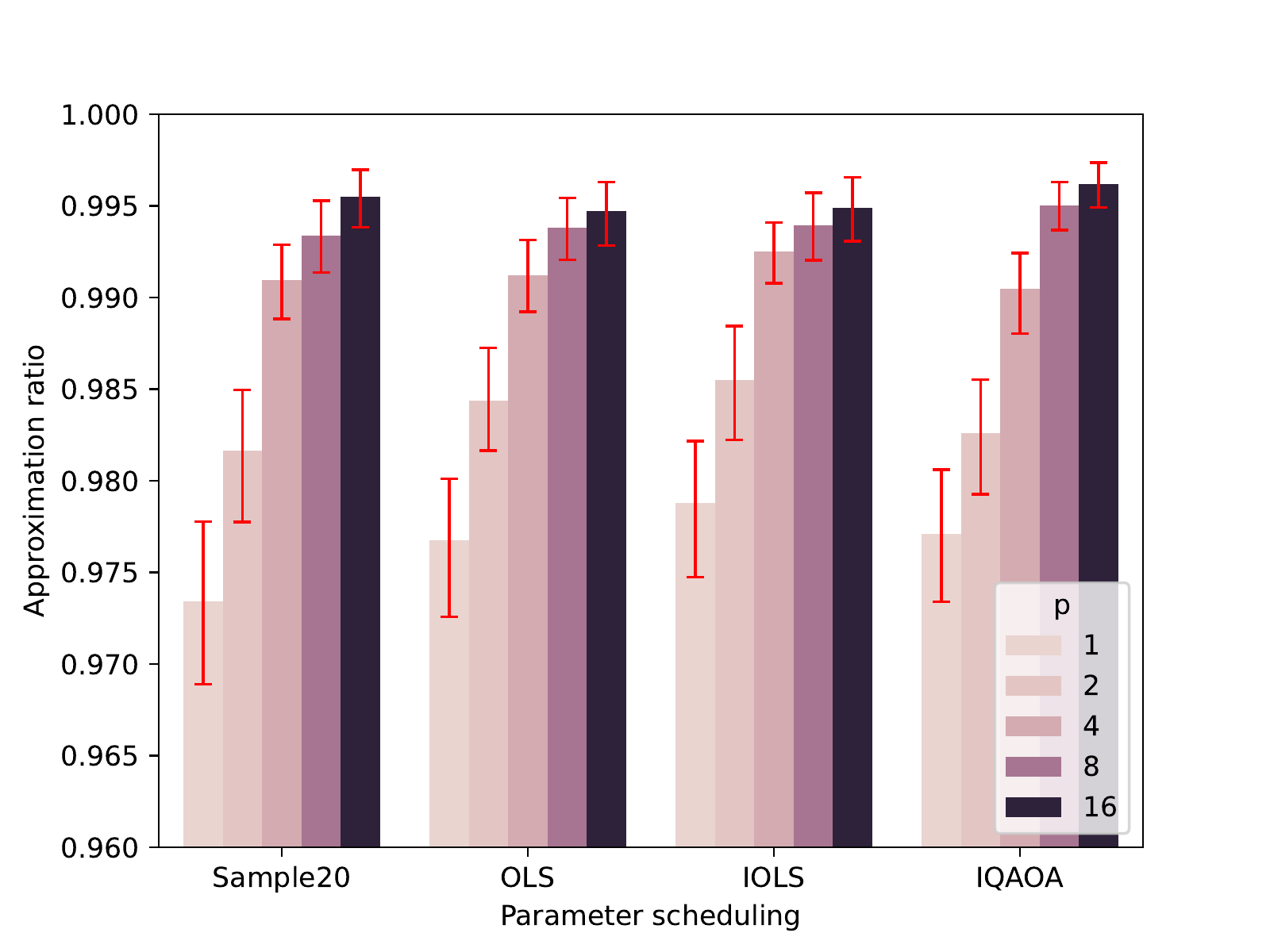}
  \caption{The approximation ratio of CVaR-QAOA in general cases.}
  \label{fig:all_cases}
\end{figure}

In Instance 2, we consider different stock pools and more general upper and lower bounds. The dataset for this experiment is taken from the Chinese Shenzhen and Shanghai Stock Exchange Securities Market. We randomly select 4-8 stocks from 4836 stocks in the market that do not contain unknown values. The precision $\alpha$ is set to 0.05, and the upper and lower bounds of each stock are randomly set between 0 and 1, but the interval between the upper and lower bounds is randomly set to 0.1, 0.2, or 0.4. Then, under the quasi-binary encoding, cases with a total qubit number exceeding 20 will be excluded. We conducted 320 experiments on each combination of four parameter scheduling methods (Sample20, OLS, IOLS, IQAOA) and five different depths ($p$=1, 2, 4, 8, 16) under the CVaR-QAOA framework. Figure \ref{fig:all_cases} shows the approximation ratio in the above experiments, ranging from 0.973 to 0.997, and the error bars still represent the 95\% confidence interval. In this instance, the average approximation ratio of the four parameter scheduling methods will increase as the depth increases, and the error range will become smaller. One speculation for this phenomenon is that the proportion of cases with few feasible solutions is large, and even if the COBYLA optimization algorithm falls into the barren plateau, it can still find the optimal solution in these cases. Therefore, when $p$ is large, its performance will not suffer from the barren plateau phenomenon. Among the four parameter scheduling methods, IQAOA still performs the best, consistent with the conclusion of instance 1.

\section{Precision Increasing Iterative Method}
\label{sec6}

Based on the simulation results of the two instances in Subsection \ref{sec::realsimulation}, we observe that the precision is too coarse to meet business expectations. No asset allocation expert would accept an investment strategy with a 50\% multiple as a proportion. However, if we increase the precision, for example, by setting $\alpha$ to one-thousandth, then the total number of qubits required in Instance 1 is 96, which already exceeds the computational limit of most quantum computers and simulators.

Therefore, in this section, we propose a precision increasing iterative method where the precision parameter $\alpha$ is exponentially reduced in each iteration, but qubits counts remain the same. The principle of this iterative method is to use QB-QAOA with a coarse precision parameter $\alpha^{(0)}$, lower bounds $l^{(0)}=l$ and upper bounds $u^{(0)}=u$ to obtain the optimal share $\mathbf{x}^{(0)}$. Then, an iteration factor $\lambda < 1$ is decided, and a range is generated around $\mathbf{x}^{(0)}$ by radiating outward according to $\lambda$, resulting in new upper and lower bounds $u^{(1)}$ and $l^{(1)}$. $\mathbf{x}^{(0)}$ should be as close to the center of $u^{(1)}$ and $l^{(1)}$ as possible, and for each asset $i$, $u^{(1)}_i-l^{(1)}_i \leq \lambda (u^{(0)}_i-l^{(0)}_i)$. Next, QB-QAOA is run again with a slightly finer precision parameter $\alpha^{(1)} = \lambda \alpha^{(0)}$ to solve the portfolio optimization problem under $u^{(1)}$ and $l^{(1)}$. In the $(k+1)$-th iteration, the precision parameter becomes $\alpha^{(k)} = \lambda^k \alpha^{(0)}$. The smaller the iteration factor $\lambda$ is, the faster the precision parameter will be reduced, but it is also more likely to fall into local optima around previous solutions if the value of $\lambda$ is small. It is recommended to choose $\lambda$ such that $1/\lambda$ is an integer, so that $u^{(k)}$ and $l^{(k)}$ are exactly integer multiples of $\alpha^{(k)}$. Otherwise, $u^{(k)}$ and $l^{(k)}$ need to be slightly adjusted to be divisible by $\alpha^{(k)}$. The algorithm of the iteration method is shown as Algorithm \ref{algo:Quasi_Binary_QAOA_iter}.

\begin{algorithm}
\caption{QuasiBinaryQAOAiter}
\label{algo:Quasi_Binary_QAOA_iter}
\begin{algorithmic}[1]
\STATE \textbf{Input:} Precision parameter $\alpha$, lower bound array $l$, upper bound array $u$, iteration factor $\lambda$, the number of iterations $interNum$, quasi-binary QAOA function QuasiBinaryQAOA.
\STATE \textbf{Output:} Shares array $x$
\STATE $x \gets$ QuasiBinaryQAOA($\alpha, l, u$)
\STATE l\_old $\gets l$
\STATE u\_old $\gets u$
\FOR{$i \gets 1$ to $interNum$}
\STATE $\lambda$\_new $\gets \lambda^i$
\STATE $\alpha$\_new $\gets \alpha \cdot \lambda$\_new
\STATE $l1 \gets x - (u - l) \cdot \lambda$\_new $/ 2$
\STATE $u1 \gets x + (u - l) \cdot \lambda$\_new $/ 2$
\STATE l\_new $\gets \max(l, l1)$
\STATE u\_new $\gets \min(u, u1)$
\STATE $x \gets$ QuasiBinaryQAOA($\lambda$\_new, $l$\_new, $u$\_new)
\ENDFOR
\RETURN $x$
\end{algorithmic}
\end{algorithm}

\begin{figure}[htbp]
  \centering
  \includegraphics[width=0.9\textwidth]{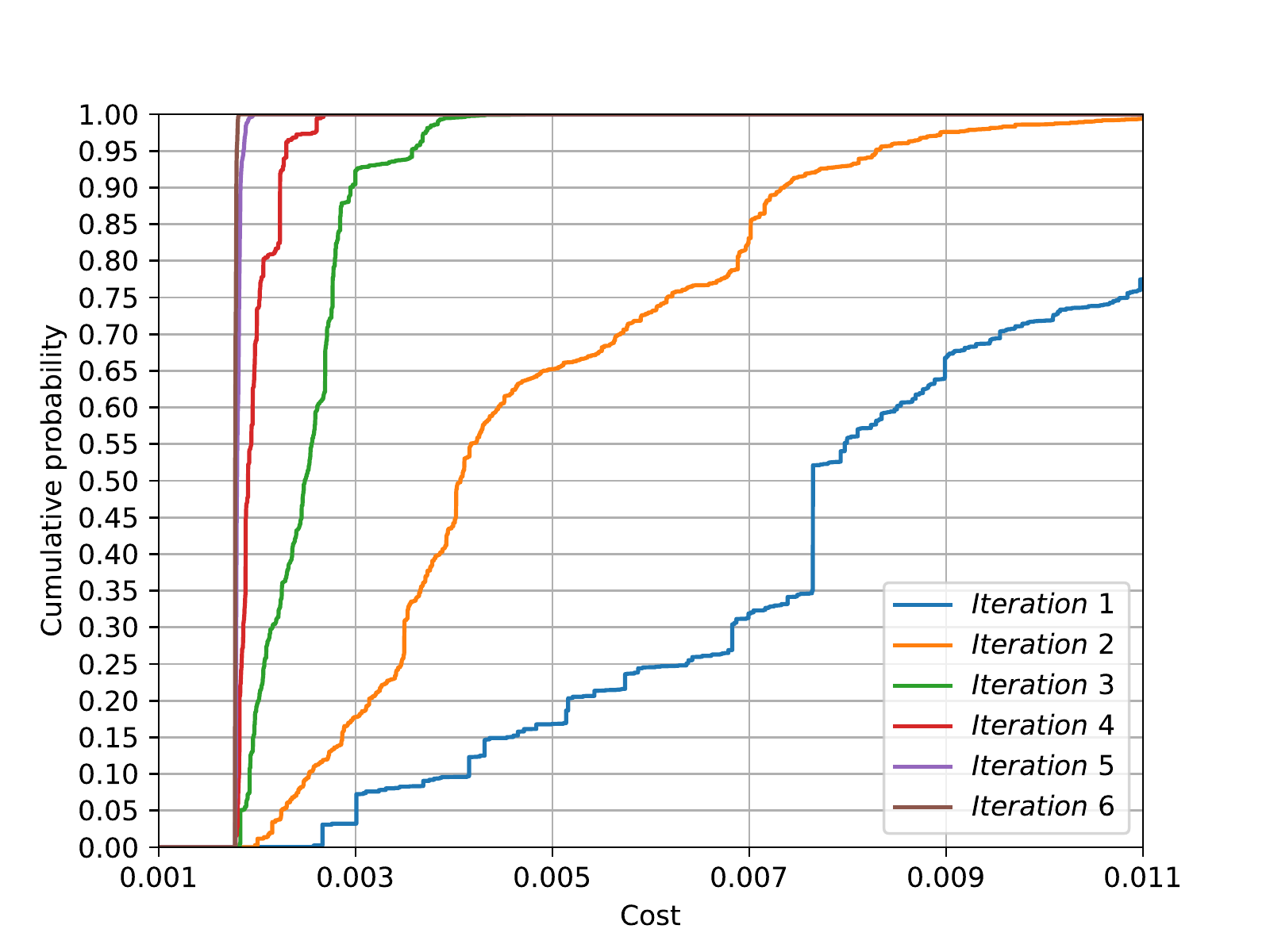}
  \caption{Cumulative distribution probability of solution costs from different precisions in precision increasing iterative method}
  \label{fig:iteration}
\end{figure}

In the simulation, we utilize the experiment parameters from Instance 1 with $p=8$ in Section \ref{sec::realsimulation}. We employ QB-QAOA with CVaR-QAOA model and IQAOA scheduling method. $\lambda$ is set to 0.5. After six iterations, we obtained the cumulative distribution probability of solution costs of six different precisions as shown in Figure \ref{fig:iteration}. It can be observed that QB-QAOA has a significant improvement in the quality of solutions under the precision increasing iterative method, and the cumulative distribution probability curve gradually converges. The worst solution in the fourth iteration is already better than the best solution in the first iteration. All six experiments used 18 qubits in the simulation.

\section{Conclusion}
\label{sec7}
In this paper, we have presented the quasi-binary based QAOA. This algorithm can be employed to solve combinatorial optimization problems with discrete constraint, bound constraint and sum constraint. The number of qubits allocated for each variable is logarithmically proportional to its range.

Moreover, we have proposed a precision increasing iterative method with QB-QAOA to solve portfolio optimization problems. In the course of iterative experiments with the same number of qubits, QB-QAOA has demonstrated a significant improvement in the quality of solutions.

Overall, our study has shown that QB-QAOA algorithm is a promising approach for solving portfolio optimization problems in a more efficient and effective manner. We hope that our findings will inspire further research in this area, and we believe that our proposed method can contribute to advancements in discrete quantum optimization algorithms.

\section*{Acknowledgments}
I would like to express my deepest gratitude and appreciation to my colleagues at CCBFT QLab, Z.Gao, X. Zhang, H. Chen, D. Yu, and H. Hou. Their valuable insights and contributions have been essential to the success of this project. I would also like to extend my sincere thanks to Q. Zhang, Z. Wang, and N. Qu at China Construction Bank Group Fintech Innovation Center for their help in verifying the phase separation Hamiltonian formulae and deriving the risk factors. Last but not least, I would like to thank K. Feng from Pengpai News for her assistance in designing the multiple images used in this paper. Without the contributions and support of these individuals, this paper would not have been possible. Thank you all once again for your invaluable assistance.

\subsection*{Funding}
This work was supported by CCB Fintech Company Limited (No. PO3522083587, No. PO3522083675) and by Chengdu Science and Technology Bureau (No. 2021-YF09-00114-GX).

\subsection*{Conflicts of Interest}
The authors have declared that there is no conflict of interest regarding the publication of this article.

\subsection*{Data Availability}
The original data analyzed in this research article were the closing prices of six well-known NASDAQ companies (Apple/Amazon/Google/Microsoft/NVIDIA/Tesla) during the period of 2022.12.02 to 2023.02.28. The data were obtained from Yahoo Finance, such as for Tesla company, the data were obtained from https://finance.yahoo.com/quote/TSLA/history?p=TSLA. The data are openly available and can be accessed freely without any restrictions from the source. The data of the Chinese stock market in Instance 2 is sourced from the stock data of the Shanghai and Shenzhen exchanges from May 4th, 2023 to August 8th, 2023, which is accessed by the open-source software package QStock.

\subsection*{Ethics approval}
This study doesn not include human or animal subjects.

\subsection*{Author contributions}
All authors contributed to the study conception and design. The first draft of the manuscript was written by Bingren Chen. The numerical simulation, the new discrete model and iterative method are presented by Bingren Chen. The construction of the quasi-binary encoding and the XYY-mixer are presented by Hanqing Wu. All authors commented on previous versions of the manuscript. All authors read and approved the final manuscript.

%Bibliography
\bibliographystyle{unsrt}  
\bibliography{ref}  

\begin{appendices}
  \section{Markowitz model}
  \label{a1}
  The Markowitz model serves the purpose of determining the proportion of funds to be invested in each asset out of the total funds, given the covariance matrix and expected returns of $N$ assets. The original Markowitz model is expressed by the following equation:

\begin{equation*}
  \label{ea}
  \tag{A}
  \begin{aligned}
  &\min_w \ w^T \mathbf{\Sigma} w \\
  &\textbf{s.t.} \  w^T \mathbf{E} \geq \mu,\ \mathbf{1}^Tw=1,
  \end{aligned}
\end{equation*}
Here, $w=(w_1, \cdots, w_N)$ represents the vector of investment proportions for the $N$ assets, $\mu$ denotes the minimum target return, $\mathbf{\Sigma}$ is the covariance matrix of returns, $\mathbf{E}$ is the vector of returns expectation, and $\mathbf{1} = (1, \cdots, 1)^T \in \mathbb{R}^N$ represents a vector of ones.

Upon performing some transformations, the Markowitz model can be expressed in an equivalent form as follows:
\begin{equation*}
  \label{eb}
  \tag{B}
  \begin{aligned}
  &\min_w \ \frac{1}{2} q w^T \Sigma w - w^T \mathbf{E}\\
  &\textbf{s.t.} \ \ \mathbf{1}^Tw=1,
  \end{aligned}
\end{equation*}
where $q \in (0, + \infty)$ represents the risk factor, where higher values indicate a more conservative investment strategy that places greater emphasis on risk, and lower values indicate a more aggressive strategy that prioritizes returns. The relationship between $q$ and $\mu$ is given by:
\begin{equation*}
  \label{ec}
  \tag{C}
  q= \frac{ac-b^2}{c\mu-b},
\end{equation*}
where the constants $a$, $b$, and $c$ are defined as:

\begin{equation*}
  \label{ed}
  \tag{D}
  \begin{aligned}
  &a = \mathbf{E}^T \Sigma^{-1} \mathbf{E},\\
  &b = \mathbf{E}^T \Sigma^{-1} I_N,\\
  &c = I_N \Sigma^{-1} I_N,
  \end{aligned}
\end{equation*}

\begin{figure}[htbp]
    \centering
    \includegraphics[width=0.9\textwidth]{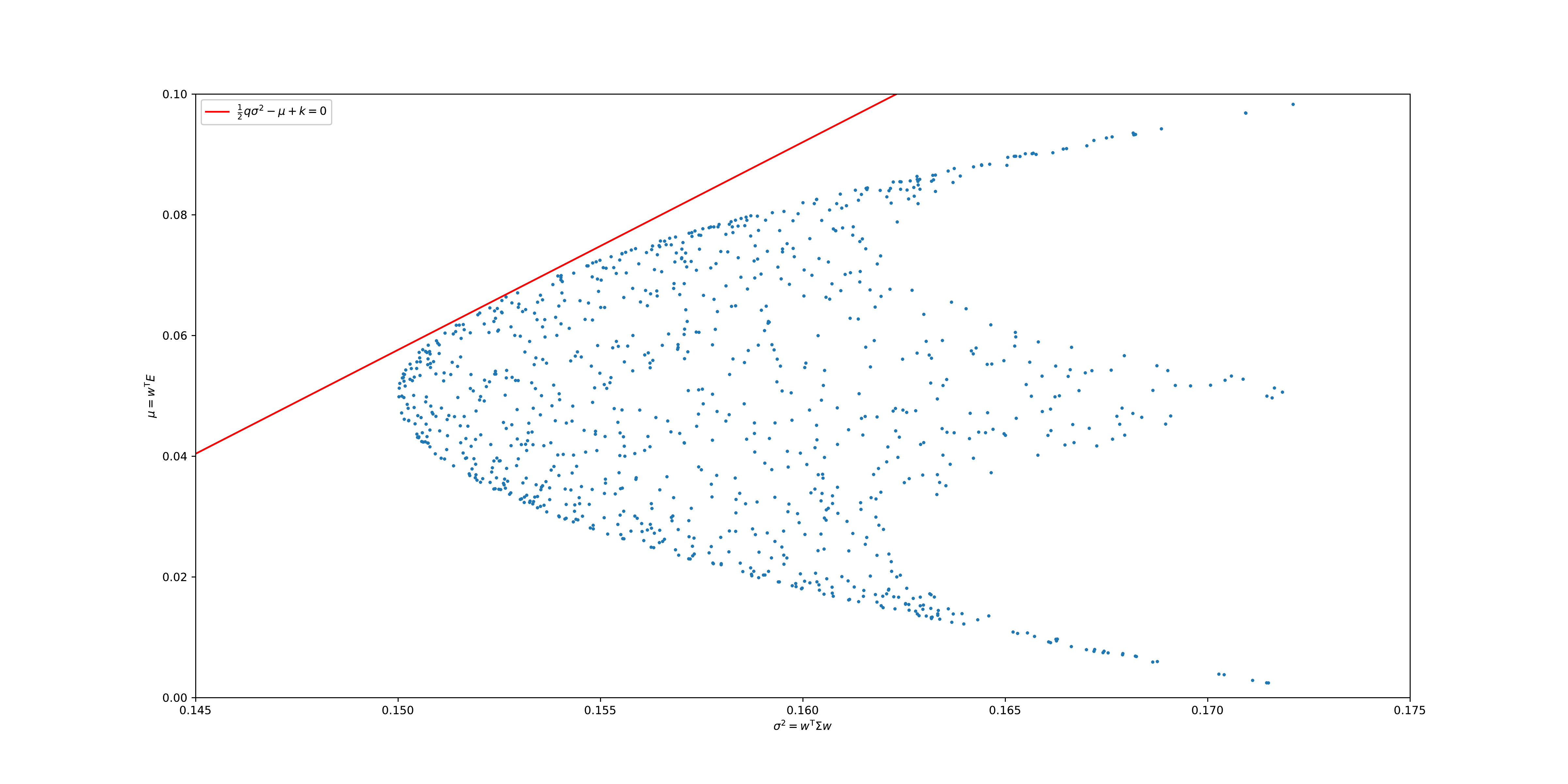}
    \caption{The efficient frontier in Markowitz model.}
    \label{fig:mw}
\end{figure}

  Figure \ref{fig:mw} illustrates the efficient frontier of the Markowitz model. The depicted scatter points correspond to Eq. \ref{ea}, which represents the values of $\sigma^2 = \min_w w^T \mathbf{\Sigma} w$ for different investment portfolios under the constraints of $w^T \mathbf{E} \geq \mu$ and $\mathbf{1}^T w = 1$. The straight line in the graph corresponds to Eq. \ref{eb}, which takes the form of $\frac{1}{2}q\sigma^2 - \mu + k = 0$ or $\sigma^2 = \frac{2(\mu-k)}{q}$, where $k$ indicates the intersection of the line with the y-axis.
  
It is worth noting that the upper boundary of the scatter points forms a curve that satisfies the constraints of $w^T \mathbf{E} = \mu$ and $\mathbf{1}^T w = 1$. By solving for $w$ and substituting it into $\sigma^2 = \min_w w^T \mathbf{\Sigma} w$, we obtain the equation of the efficient frontier curve as:

\begin{equation*}
\label{ee}
\tag{E}
\sigma^2 = \frac{a-2b\mu+c\mu^2}{ac-b^2},
\end{equation*}

where the values of $a$, $b$, and $c$ are presented in Eq. \ref{ed}.

We should expect that the optimal solutions obtained from Eq. \ref{ea} and Eq. \ref{eb} would coincide with each other. Therefore, the optimal expectation $\mu$ and risk $\sigma^2$ should satisfy both the equation of the straight line and the equation of the curve.

Now, we calculate the partial derivative of Eq. \ref{ee} with respect to $\mu$, yielding:

\begin{equation*}
\label{ef}
\tag{F}
\frac{\partial \sigma^2}{\partial \mu}=\frac{-2b+2c\mu}{ac-b^2}.
\end{equation*}

The value of Eq. \ref{ef} should be equal to the slope of the straight line, which is $\frac{2}{q}$. Hence, by solving:

\begin{equation*}
\label{eg}
\tag{G}
\frac{2}{q}=\frac{-2b+2c\mu}{ac-b^2},
\end{equation*}

we can obtain the modified equation as Eq. \ref{ec}. When there are other constraints in Eq. \ref{ea} and Eq. \ref{eb}, Eq. \ref{ec} cannot guarantee that the two are equivalent, but we can still use it to choose the value of $q$ as a reference.

\end{appendices}

\end{document}